\documentclass[aps,prb,twocolumn,longbibliography,superscriptaddress]{revtex4-1}
\usepackage{epsfig}
\usepackage{epstopdf}
\usepackage{amsmath}
\usepackage{amsfonts}
\usepackage{amssymb}
\usepackage{hyperref}
\usepackage{bm}
\usepackage{makecell}
\usepackage{rotating}
\usepackage{hyperref}

\usepackage{graphicx}
\usepackage{dcolumn}
\usepackage{bm}
\usepackage{color}

\usepackage{tikz,xcolor,hyperref}

\definecolor{lime}{HTML}{A6CE39}
\DeclareRobustCommand{\orcidicon}{%
	\begin{tikzpicture}
	\draw[lime, fill=lime] (0,0)
	circle [radius=0.16]
	node[white] {{\fontfamily{qag}\selectfont \tiny ID}};
	\draw[white, fill=white] (-0.0625,0.095)
	circle [radius=0.007];
	\end{tikzpicture}
	\hspace{-2mm}
}

\foreach \x in {A, ..., Z}{%
	\expandafter\xdef\csname orcid\x\endcsname{\noexpand\href{https://orcid.org/\csname orcidauthor\x\endcsname}{\noexpand\orcidicon}}
}

\begin{document}

\title{
Momentum-resolved spin splitting in Mn-doped trivial CdTe \\ and topological HgTe semiconductors
}

\author{Carmine Autieri\orcidA}\email{autieri@MagTop.ifpan.edu.pl}
\affiliation{International Research Centre MagTop, Institute of Physics, Polish Academy of Sciences, Aleja Lotnikow 32/46, PL-02668 Warsaw, Poland}
\affiliation{Consiglio Nazionale delle Ricerche CNR-SPIN, UOS Salerno, 84084 Fisciano (Salerno), Italy}

\author{Cezary \'{S}liwa\orcidC}
\affiliation{Institute of Physics, Polish Academy of Sciences, Aleja Lotnikow 32/46, PL-02668 Warsaw, Poland}

\author{Rajibul Islam\orcidD}
\affiliation{International Research Centre MagTop, Institute of Physics, Polish Academy of Sciences, Aleja Lotnikow 32/46, PL-02668 Warsaw, Poland}

\author{Giuseppe Cuono\orcidE}
\affiliation{International Research Centre MagTop, Institute of Physics, Polish Academy of Sciences, Aleja Lotnikow 32/46, PL-02668 Warsaw, Poland}

\author{Tomasz Dietl\orcidB}
\affiliation{International Research Centre MagTop, Institute of Physics, Polish Academy of Sciences, Aleja Lotnikow 32/46, PL-02668 Warsaw, Poland}
\affiliation{WPI-Advanced Institute for Materials Research, Tohoku University, Sendai 980-8577, Japan}


\begin{abstract}
Exchange coupling between localized spins and band or topological states accounts for giant magnetotransport and magnetooptical effects as well as determines spin-spin interactions in magnetic insulators and semiconductors. However, even in archetypical dilute magnetic semiconductors such as Cd$_{1-x}$Mn$_x$Te and Hg$_{1-x}$Mn$_x$Te the evolution of this coupling with the wave vector ${\bm{k}}$ is not understood. For instance, a series of experiments have demonstrated that exchange-induced splitting of magnetooptical spectra of Cd$_{1-x}$Mn$_x$Te and Zn$_{1-x}$Mn$_x$Te at the $L$ points of the Brillouin zone is, in contradiction to the existing theories, more than one order of magnitude smaller compared to its value at the zone center and can show an unexpected sign of the effective Land\'e factors, opposite to that found for topological Hg$_{1-x}$Mn$_x$Te. The origin of these findings we elucidate quantitatively by combining: (i) relativistic first-principles density functional calculations with the modified Becke-Johnson exchange-correlation potential; (ii) a tight-binding approach that takes carefully into account ${\bm k}$-dependence of the potential and kinetic $sp$-$d$ exchange interactions; (iii) a theory of magnetic circular dichroism (MCD) for $E_1$ and $E_1$ + $\Delta_1$ optical transitions, developed here within the envelope function $k \cdot p$ formalism for the $L$ point of the Brillouin zone in zinc-blende crystals. This combination of methods leads to the conclusion that the physics of MCD at the {\em boundary} of the Brillouin zone is strongly affected by the strength of two relativistic effects in particular compounds: (i) the mass-velocity term that controls the distance of the conduction band at the $L$ point to the upper Hubbard $d^6$ band of Mn ions and, thus, a relative magnitude and sign of the exchange splittings in the conduction and valence bands; (ii) the spin-momentum locking by spin-orbit coupling that reduces exchange splitting depending on the orientation of particular $L$ valleys with respect to the magnetization direction.

\end{abstract}


\maketitle

\section{Introduction}
Dilute magnetic semiconductors (DMSs), such as Cd$_{1-x}$Mn$_x$Te and Hg$_{1-x}$Mn$_x$Te, have played a central role in the demonstrating and describing a strong and intricate influence of the $sp$-$d$ exchange interactions upon effective mass states in semiconductors \cite{Kossut:2010_B,Dietl:1994_B,Furdyna:1988_JAP}, paving the way for the rise of dilute ferromagnetic semiconductors \cite{Dietl:2014_RMP} and magnetic topological insulators \cite{Ke:2018_ARCMP,Tokura:2019_NRP}. One of the key characteristics of DMSs is a giant spin splitting of bands proportional to the field-induced and temperature-dependent magnetization of paramagnetic Mn$^{2+}$ ions, ${\bm{M}}(T,{\bm{H}})$. In the case of high electron mobility modulation-doped Cd$_{1-x}$Mn$_x$Te/Cd$_{1-y}$Mg$_y$Te  heterostructures, the exchange splitting leads to crossings of spin-resolved Landau levels, at which the quantum Hall ferromagnet forms at low temperatures \cite{Jaroszynski:2002_PRL}. It has been recently proposed that magnetic domains of this ferromagnet, if proximitized by a superconductor, can host Majorana modes \cite{Kazakov:2016_PRB,Kazakov:2017_PRL,Simion:2018_PRB}.  Similarly, Hg$_{1-x}$Mn$_x$Te/Hg$_{1-y}$Cd$_y$Te quantum wells of an appropriate thickness and Mn cation concentration $x\lesssim 7$\%, which ensures the inverted band structure, may show a ladder of spin sublevels in the magnetic field, enabling the appearance of the quantum anomalous Hall effect \cite{Liu:2008_PRLb}. Interestingly,  Hg$_{1-x}$Mn$_x$Te samples in either a bulk \cite{Sawicki:1983_P} or quantum well \cite{Shamim:2020_SA} form show, in the vicinity of the topological phase transition, magnetotransport signatures of quantized Landau levels already below 50 mT, if the Fermi level is tuned to the Dirac point. It has also been predicted that biaxial tensile strain will turn the topological semimetal Hg$_{1-x-y}$Cd$_x$Mn$_y$Te into Weyl's semimetal for non-zero magnetization of Mn spins \cite{Bulmash:2014_PRB}.

According to the present insight there are two exchange mechanisms involved in the interaction ${\cal{H}}_{sp-d} = -J{\bm{s}}\cdot{\bm{S}}$ between effective mass electrons in the vicinity of $\Gamma$ and localized spins residing on the half-filled Mn$^{2+}$ $d$-shells \cite{Kacman:2001_SST}. The first on them is  ferromagnetic direct (potential) exchange $J_{sd}$ between band carriers with wave functions derived from Mn $s$ orbitals and electrons localized on the open Mn $d$ shells, usually denoted $N_0\alpha$, typically of the order of 0.2\,eV.  The second one is the antiferromagnetic kinetic exchange between band carriers with anion $p$-type wave functions and $d$ electrons, of the order of $J_{pd} \equiv N_0\beta \approx -1$\,eV.  Incorporation of these interactions into an appropriate multi-band $k \cdot p$ Hamiltonian allows one to describe satisfactorily various spectacular magnetotransport and magnetooptical phenomena for carriers near the $\Gamma$ point of the Brillouin zone as a function of ${\bm{M}}(T,{\bm{H}})$ \cite{Kossut:2010_B,Dietl:1994_B,Furdyna:1988_JAP,Bastard:1978_JdP}, particularly if effects of strong coupling are taken into account \cite{Dietl:2008_PRB}.

However, in contrast to the $\Gamma$ point, the physics of exchange splittings at the $L$ points of the Brillouin zone is challenging: a series of magnetoreflectivity and magnetic circular dichroism (MCD) studies, notably for Cd$_{1-x}$Mn$_x$Te \cite{Ginter:1983_SSC,Coquillat:1986_SSC,Ando:2011_JAP}, has revealed that the magnitudes of spectra splittings for two circular light polarizations at the $L$ points ($E_1$ and $E_1 +\Delta_1$ transitions) are smaller by a factor of about sixteen compared to the value at the $\Gamma$ point, an effect not explained by tight-binding modeling  \cite{Ginter:1983_SSC,Bhattacharjee:1990_PRB}. Furthermore, effective Land\'e factors corresponding to these transitions can show an unexpected sign \cite{Ando:2011_JAP}. The situation is also unsettled in Hg$_{1-x}$Mn$_x$Te, in which a large magnitude of spin-orbit-driven spin-splittings accounts for a controversy concerning the actual values of the $sp$-$d$ exchange integrals at the $\Gamma$ point \cite{Dietl:1994_B}, making their comparison to spin-splitting values at the $L$ points \cite{Coquillat:1989_PRB} not conclusive. Accordingly, it has been pointed out that the electronic structures of II-VI DMSs have not been as well clarified as we previously believed \cite{Ando:2011_JAP}. Among other issues, this fact may preclude a meaningful evaluation of the role played by interband spin polarization in mediating indirect exchange interactions between magnetic ions. This Bloembergen-Rowland mechanism \cite{Bloembergen:1955_PR} is known to play a sizable role in \emph{p}-type dilute ferromagnetic semiconductors, in which it involves virtual transitions between hole valence subbands \cite{Dietl:2001_PRB,Ferrand:2001_PRB}. Moreover, this spin-spin exchange is expected to be particularly important in the absence of carriers in the inverted band structure case (such as Hg$_{1-x}$Mn$_x$Te), in which both the valence and conduction bands are primarily built of anion $p$-type wave functions \cite{Bastard:1979_PRB,Yu:2010_S}.

In the last years, several {\em ab initio} studies of Cd$_{1-x}$Mn$_x$Te have been carried out \cite{Wei:1987_PRB,Larson:1988_PRB,Merad:2006_JMMM,Liu:2008_PSSB,Echeverria:2009_PRB,Verma:2011_JMMM,Wua:2015_CMS,Linneweber:2017_PRB}.
However, these works have not attempted to elucidate the origin of the anomalously exchange-induced splittings of optical spectra corresponding to transitions at the Brillouin zone boundary.
In our work, we determine ${\bm{k}}$-dependent exchange  splittings of bands for both Cd$_{1-x}$Mn$_x$Te and Hg$_{1-x}$Mn$_x$Te employing the density functional theory (DFT), the tight-binding approximation (TBA), and the $k \cdot p$ envelope function formalism sequentially.
Our quantitative results demonstrate that competition between ferromagnetic and antiferromagnetic exchange interactions, the relativistic mass-velocity term, and the spin-momentum locking by spin-orbit coupling constitute the essential ingredients determining magnitudes of spectral splittings at the $L$ points of the Brillouin zone in Cd$_{1-x}$Mn$_x$Te and Hg$_{1-x}$Mn$_x$Te,  hitherto regarded as not understood \cite{Ginter:1983_SSC,Coquillat:1986_SSC,Ando:2011_JAP,Bhattacharjee:1990_PRB,Coquillat:1989_PRB}. Important outcomes of our work are also: the minimal tight-binding model that describes the electronic band structure of CdTe, HgTe, Cd$_{1-x}$Mn$_x$Te, and Hg$_{1-x}$Mn$_x$Te in the whole Brillouin zone quantitatively, and the $k \cdot p$ Hamiltonian suitable for modeling phenomena involving $L$-valleys in compounds  with a zinc-blende crystal structure.

\section{Computational methodology}
\label{sec:methodology}

\subsection{Overview of theoretical approach}
We aim at the determination of exchange splittings in the whole Brilloiun zone and then of MCD spectra in two classes of DMSs as well as at the elucidation of the origin of a substantial reduction of MCD at the $L$ point compared to the zone center.  This program requires consideration of spin-orbit and sp-d exchange splittings on equal footing. Furthermore, we would like to obtain a minimal tight-binding (TB) model suitable for the description of phenomena, such as spin-spin interactions, which depend on the band structure in the whole Brillouin zone. It may appear that the accomplishment of such goals is straightforward by modern fully relativistic DFT implementations, notably, employing approaches developed for alloys, such as the special quasi-random structure (SQS)  \cite{Zunger:1990_PRL}.  Surprisingly, however, we have encountered several challenges.

First, as shown in Secs.~IIIA and IIIB, by using generalized gradient approximation (GGA) with intra-site Hubbard $U$ for Mn $d$ electrons, we have been able to obtain information about main effects leading to a strong dependence of exchange band splittings on the ${\bm{k}}$-vector without spin-orbit coupling (SOC). At the same time, however, our findings confirm that the use of this computationally effective method may lead to qualitatively misleading information in DMS \cite{Zunger:2010_P}. Indeed, the GGA + U provides not only wrong values of energy gaps, but also of exchange energies whose values are rather sensitive to the distance of Mn d-levels and bands, particularly away from the $\Gamma$ point. To overcome this difficulty, we have implemented the modified Becke-Johnson exchange-correlation potential (MBJLDA) \cite{Becke:2006_JCP,Tran:2009_PRL} (Se.~IIIC), whose use is, however,  more computationally demanding,  particularly within SQS.

Second, with our present expertise and computation resources, we have been unable to find an effective unfolding procedure that would allow us to tell band spilittings originating from exchange interactions, SOC, and  band folding associated with a finite supercell size, particularly taking into account that both exchange and spin-orbit splittings depend on the ${\bm{k}}$-vector and magnetization directions. By contrast, the  MBJLDA provides band structure and energy gaps in good agreement with experimental values for both CdTe and HgTe as well as proper positions of Mn levels in respect to bands.  Accordingly, we have used MBJLDA information to obtain a versatile TB model (Sec.~IIID), to which $sp$-$d$ exchange interactions can readily be incorporated within the Schrieffer-Wolf procedure \cite{Schrieffer:1966_PR} (Sec.~IIIE). In this way, we obtain a tool for determining the magnitudes and signs of band splittings for any ${\bm{k}}$-vector and magnetization values and directions.

Third, there is an on-going discussion (which we recall in Sec.~IIIF) on how to determine optical matrix elements within TB approaches. There is no such ambiguity within the $\bm{k}\cdot\bm{p}$ envelope function method. We have, therefore,  developed this approach for the $L$-point of Brillouin zone in zinc-blende semiconductors (Appendix), which has allowed us, together with energy values from the TB data, to determined optical splittings of the MCD lines with no adjustable parameters (Sec.~IIIG).

\subsection{Computation details}
We have performed first-principles DFT
calculations by using the relativistic VASP package based
on plane-wave basis set and projector augmented wave method \cite{Kresse:1996_CMS}.
We perform a fully relativistic calculation for the core-electrons while the valence electrons are treated in a scalar approximation considering the mass-velocity and the Darwin terms.
Spin-orbit coupling of the valence electrons is included using the second-variation method and the scalar-relativistic eigenfunctions of the valence states \cite{Hafner:2008_JCC}.

A plane-wave  energy cut-off of 400~eV has been used. For the bulk, we have performed the calculations using 8$\times$8$\times$8 $\bm{k}$-point Monkhorst-Pack grid with 176 $k$-points in the absence of SOC and with 512 $k$-points in the presence of SOC
in the irreducible Brillouin zone.
We use the experimental lattice constants corresponding to $a_0 =6.46152$\,{\AA} for HgTe and 6.4815\,{\AA} for CdTe \cite{Skauli:2001_JCG}.

For the treatment of exchange-correlation, either Perdew-Burke-Ernzerhof (PBE)
GGA\cite{Perdew:1996_PRL} or the  MBJLDA\cite{Becke:2006_JCP,Tran:2009_PRL} have been applied.
According to the computed band structures in GGA, the magnitudes of the bandgap $E_0 = E(\Gamma_6) - E(\Gamma_8)$ are $0.77$\,eV and $-0.50$\,eV for CdTe and HgTe, to be compared to experimental values at 4.2\,K $E_0 = 1.60$\,eV and $-0.30$\,eV, respectively. These discrepancies reflect the well-known inaccuracies of the GGA in the evaluation of the bandgap. Thus, to improve the tight-binding parametrization of CdTe and HgTe band structures, the MBJLDA has been employed to determine the hopping parameters.
Our results, obtained within this computationally more demanding approach, confirm that the determined magnitudes of the band gaps \cite{Camargo:2012_PRB}, as well as of spin-orbit splittings, are close to experimental values.

The effect of Mn doping in Cd$_{1-x}$Mn$_{x}$Te and Hg$_{1-x}$Mn$_{x}$Te has been studied using a $4\times4\times4$ supercell with  64 anions and 64 cations. We use the SQS \cite{Zunger:1990_PRL} to model the distribution of cation-substitutional Mn atoms in the supercell. To create a large SQS model, we used the mcsqs algorithm \cite{Walle:2013_Calphad} within the framework of alloy theoretic automated toolkit (ATAT) \cite{Walle:2002_Calphad}.
The mcsqs method is based on the Monte Carlo simulated annealing loop with an objective function that searches for a perfectly matching maximum number of correlation functions for a fixed shape of the supercell along with the occupation of the atomic site by minimizing the objective function.
The doublet, triplet, and quadruplet clusters are generated using the cordump utility of the ATAT toolkit. We use the parameters $-2$, $-3$, and $-4$ for which the longest pair, triplet, and quadruplet correlation distance to be matched is 2.0, 1.5, and 1.0 lattice constants, respectively. To create the best SQS structure, we produce all possible structures and choose that for which the correlation difference with respect to a random structure is closest to zero.
The SQS calculations have been done using a 2$\times$2$\times$2 $\bm{k}$-point grid.
For numerical efficiency, we have used the SQS just in combination with the GGA.

\begin{figure*}[tb]
\includegraphics[width=6.2cm,angle=270]{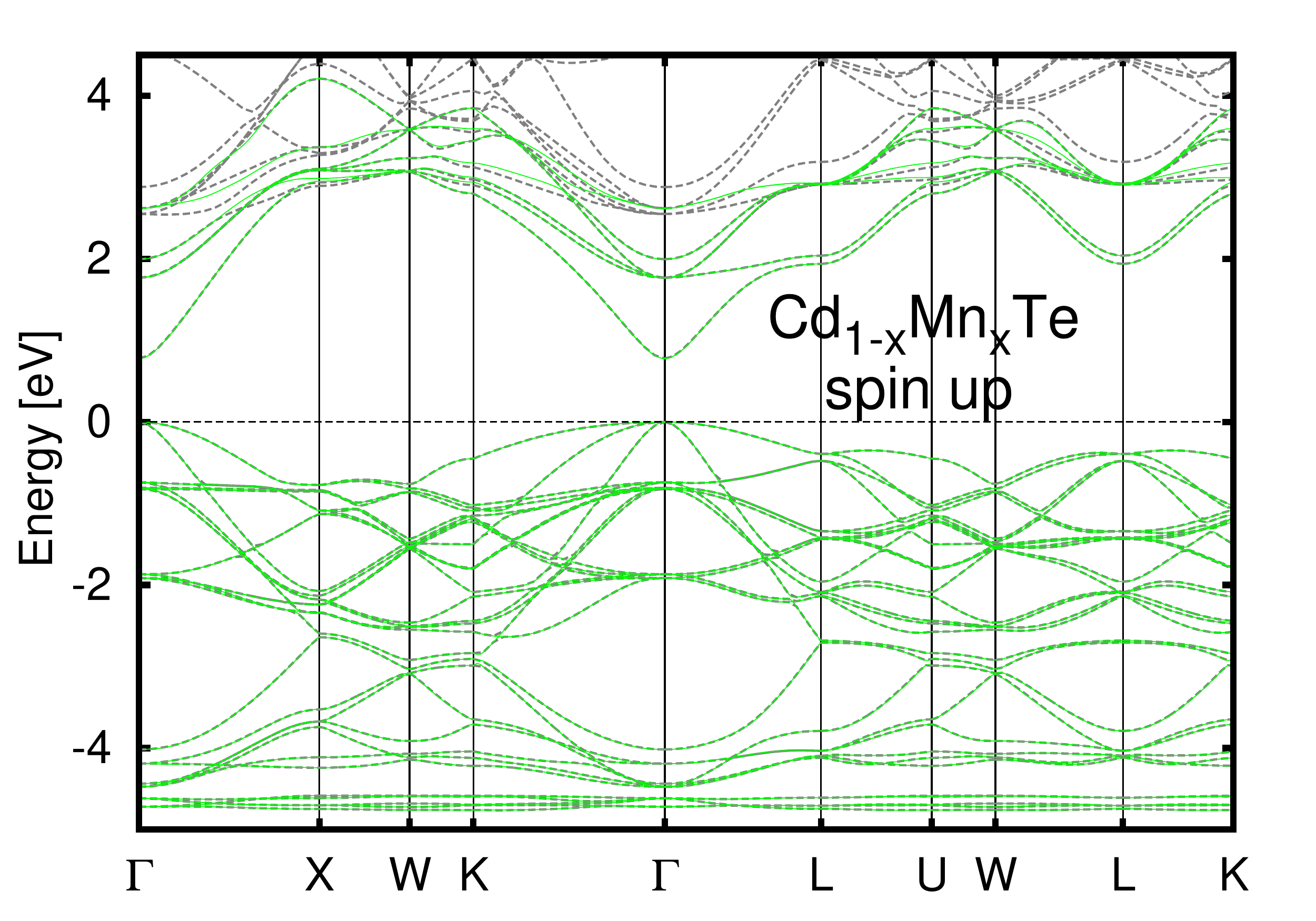}
\includegraphics[width=6.2cm,angle=270]{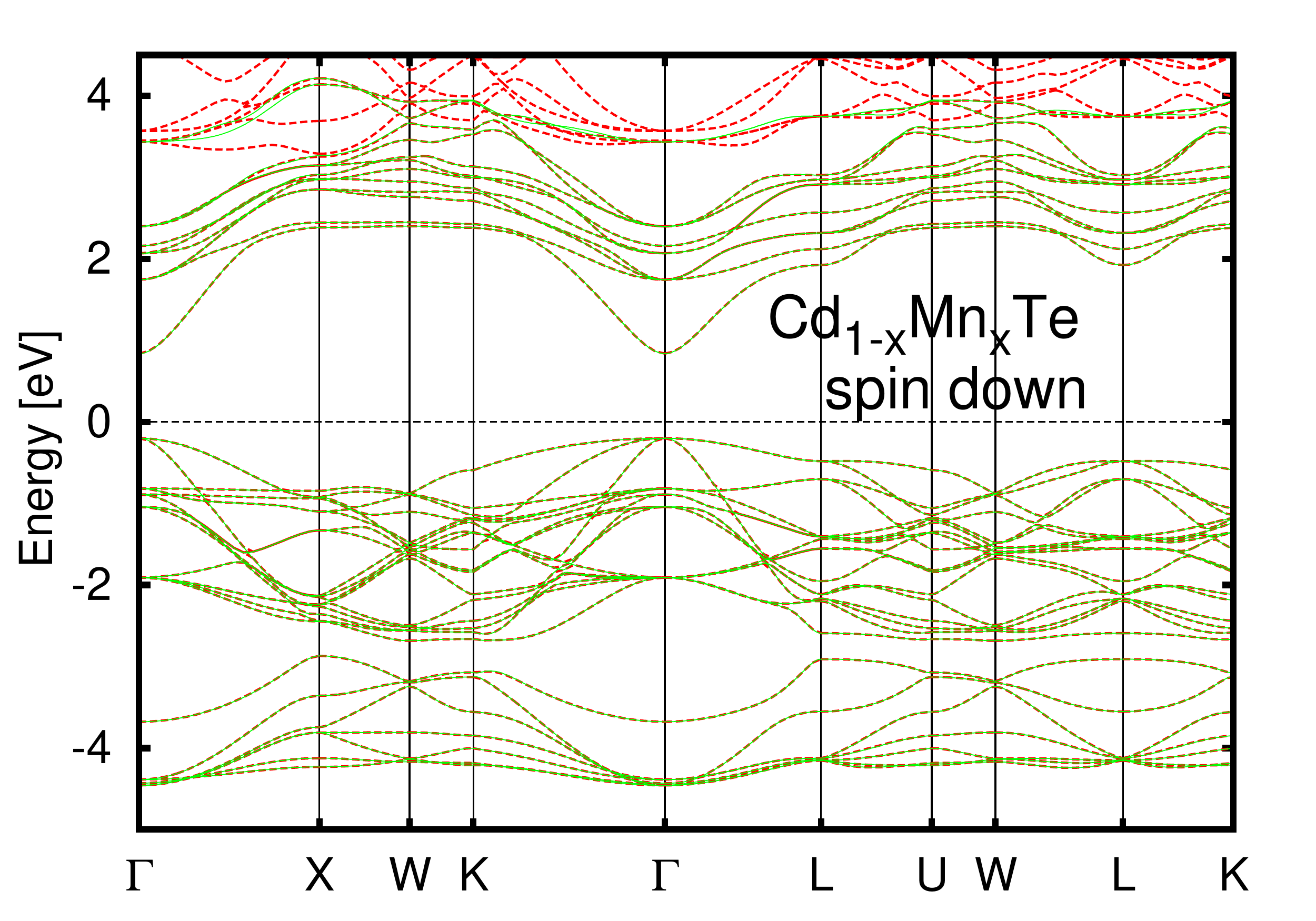}
\caption{GGA band structure obtained using the MLWF for Cd$_{0.875}$Mn$_{0.125}$Te, $U_{\text{Mn}}=5$\,eV,
and assuming ferromagnetic alinement of Mn spins without spin-orbit coupling.
A  grey (red) dashed line in the left (right) panel depicts the spin up (down) channel.
The interpolated Cd-$s$, Te-$p$, and Mn-$d$ Wannier bands are shown by a solid green line.
Zero energy is set at
the valence band top.}
\label{fig:CdMnTe_bands}
\end{figure*}

In our work, we focused on Mn content $x=2/64, 4/64$, and $8/64$. Since we look for magnitudes of $sp$-$d$ exchange splittings,  the Mn magnetic moments are always ferromagnetically aligned.
The Hubbard $U$ effects for Mn open $d$ shell have been included. We use the values of $U_{\text{Mn}}=3$, 5 and 7\,eV \cite{Paul:2014_JAC,Autieri:2014_NJP,Keshavarz:2017_PRB} and
$J_{\text{H}} = 0.15U$\,eV for the Mn-$3d$ states.
After obtaining the Bloch wave functions in density functional theory, the maximally localized Wannier functions \cite{Marzari:1997_PRB,Souza:2001_PRB} (MLWF)
are constructed using the WANNIER90 code \cite{Mostofi:2008_CPC}.  We used the Slater-Koster interpolation scheme based on Wannier functions to extract
electronic bands' character at low energies.

Quantities of interest here are effective exchange energies $J_c({\bm{k}})$ and $J_v({\bm{k}})$ calculated from ${\bm{k}}$-dependent splittings of the lowest conduction and highest valence bands, generated by exchange interactions with Mn spins $S = 5/2$, aligned by an external magnetic field,
\begin{equation}
J_c(\bm{k})=\frac{{\Delta}E_c}{xS}=\frac{E_c^\downarrow-E_c^\uparrow}{xS}, \quad
J_v(\bm{k})=\frac{{\Delta}E_v}{xS}=\frac{E_v^\downarrow-E_v^\uparrow}{xS}.
\label{eq: J_cJ_v}
\end{equation}
According to this definition, in the weak coupling limit and for the normal band ordering, i.e., for Cd$_{1-x}$Mn$_x$Te, $J_c(k =0) \equiv N_0\alpha$ and $J_v(k =0) \equiv N_0\beta$, where $N_0$ is the cation concentration, whereas $\alpha$ and $\beta$ are $s$-$d$ and $p$-$d$ exchange integrals according to the DMS  literature \cite{Kacman:2001_SST,Wei:1987_PRB,Larson:1988_PRB}. The same situation takes place in the case of Hg$_{1-x}$Mn$_x$Te with $x \gtrsim 0.07$ \cite{Furdyna:1988_JAP}. However, at lower $x$, Hg$_{1-x}$Mn$_x$Te is a zero-gap semiconductor with an inverted band structure (topological zero-gap semiconductor) for which the $s$-type $\Gamma_6$ band is below the $\Gamma_8$ $j=3/2$ multiplet forming the conduction and valence bands. In this case, we consider the spin-splitting of the $\Gamma_6$ band below the Fermi level as $J_c$. We note also that because of antiferromagnetic interactions between Mn spins, an effective Mn concentration $x_{\text{eff}}$ that contributes to the $sp-d$ exchange splitting of bands in a magnetic field is much smaller than $x$, typically $x_{\text{eff}} \lesssim 5$\% for any $x$ in relevant magnetic fields $\mu_0H \lesssim 6$\,T \cite{Gaj:1979_SSC}. For random distribution of Mn over cation sites, these antiferromagnetic interactions result in spin-glass freezing at low temperatures \cite{Galazka:1980_PRB,Mycielski:1984_SSC}.

\section{Results}
\subsection{GGA band structure for Cd$_{1-x}$Mn$_{x}$Te and Hg$_{1-x}$Mn$_{x}$Te without spin-orbit coupling}

We discuss first the electronic structure of  Cd$_{1-x}$Mn$_{x}$Te and Hg$_{1-x}$Mn$_{x}$Te computed with relativistic corrections in the scalar approximation, i.e.,  taking into account the Darwin and mass-velocity terms (essential in Hg$_{1-x}$Mn$_{x}$Te) but neglecting SOC. Such an approach allows us to extract spin splittings solely due to the exchange interactions between host and Mn spins, i.e., effective exchange integrals $J_c$ and $J_v$ for relevant bands and arbitrary $\bm{k}$-vectors in the Brillouin zone.

Figure \ref{fig:CdMnTe_bands} presents the electronic structure of Cd$_{0.875}$Mn$_{0.125}$Te for spin up and spin down evaluated assuming $U_{\text{Mn}}=5$\,eV. The Mn lower and upper Hubbard $3d$-bands reside around 4.6\,eV below and 2.5\,eV above the valence band top, respectively.
Hence, in agreement with photoelectron spectroscopy \cite{Mizokawa:2002_PRB}, an effective Hubbard energy of Mn-$3d$ electrons is 7.1\,eV for $U_{\text{Mn}}=5$\,eV and, of course,  would increase with the increasing $U_{\text{Mn}}$. At the same time, experimental data \cite{Mizokawa:2002_PRB} indicate that the Mn $d$-bands reside by about 1\,eV higher with respect to host bands than implied by our DFT results.
In the whole Brillouin zone and for both spin channels, the lowest unoccupied states consist mainly of Cd-$5s$ states, whereas Te-$5p$ states give a dominant contribution to the highest occupied bands. To estimate the orbital contribution in DFT, we evaluate the system
at the $\Gamma$ point, where the $s$-states are decoupled from the $p$ and $d$-states.
For the (Mn,Cd)Te, the conduction band is composed roughly by 70\% Cd-$s$ states and 30\% Te-$s$ states with a minor contribution from the impurity Mn-$s$ states for the low Mn concentration $x$ in question. The conduction band at the $\Gamma$ point is composed roughly by 80\% Te-$p$ states and 20\% Cd-$d$ states with a minor contribution from the impurity Mn-d states.

\begin{figure}[tb]
\centering
\includegraphics[width=6.4cm,angle=270]{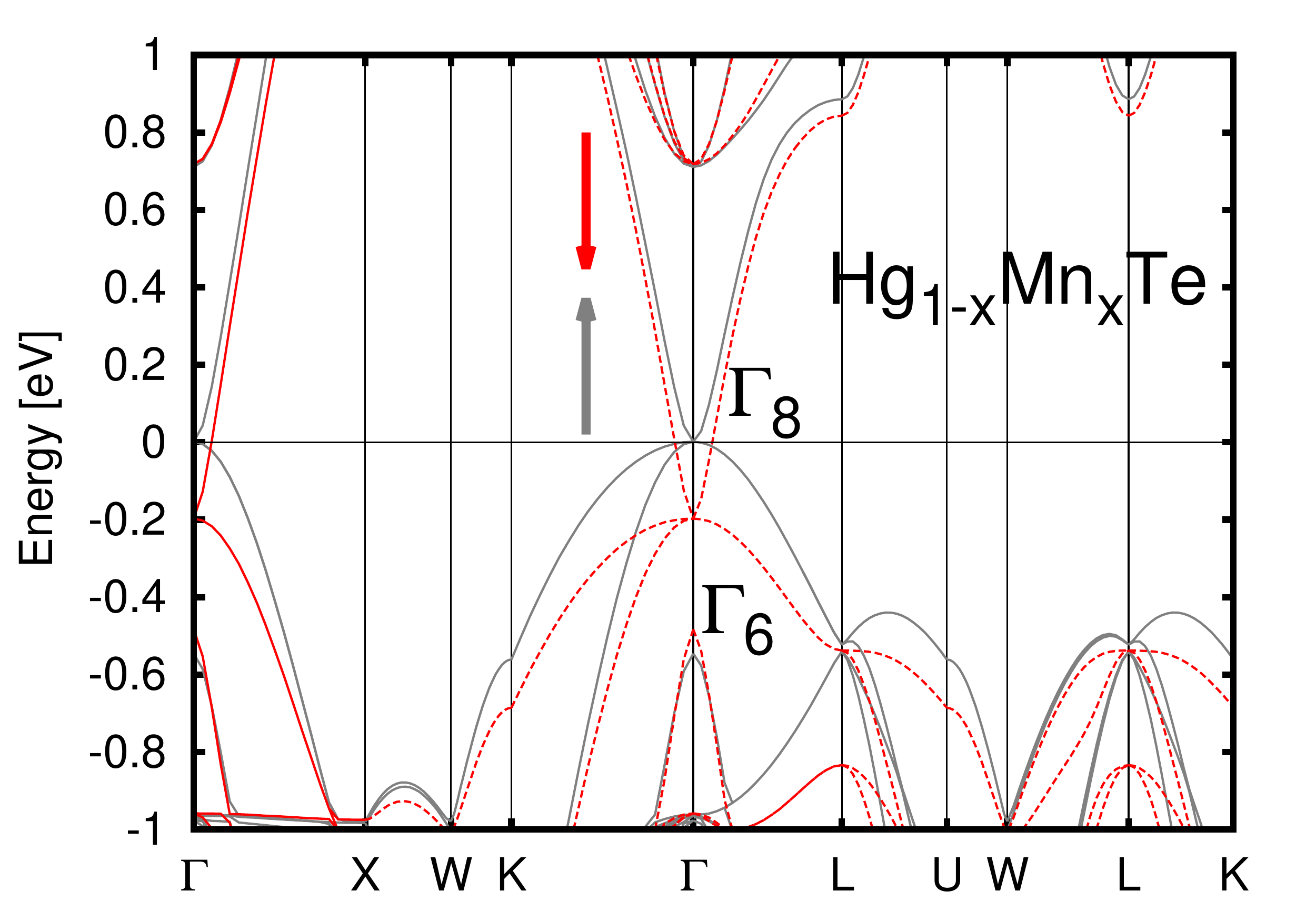}
\caption{GGA band structure of Hg$_{0.875}$Mn$_{0.125}$Te with ferromagnetically aligned Mn spins
without spin-orbit coupling and for $U_{\text{Mn}}=5$\,eV. Zero energy is set at
the valence band top.
Bands for spin up and spin down with respect to Mn spins' direction are shown by solid grey and red dash lines, respectively. }
\label{fig:HgMnTebands}
\end{figure}

From the electronic structure of Hg$_{0.875}$Mn$_{0.125}$Te at $U_{\text{Mn}}= 5$\,eV without SOC,
the effective Hubbard energy of Mn-$3d$ electrons at the $\Gamma$ point is 7.8\,eV for $U_{\text{Mn}}=5$\,eV.
As shown in Fig.~\ref{fig:HgMnTebands}, the $\Gamma_6$ and $\Gamma_8$ bands are inverted in Hg$_{1-x}$Mn$_x$Te, resulting in a topological character of the compound. The relativistic Darwin term gives a weak positive contribution to the energy of the $s$-bands in heavy atoms like Hg. In contrast, the relativistic mass-velocity term provides a strong negative energy shift, accounting for the band inversion. We can see in Fig.~\ref{fig:HgMnTebands} that the $\Gamma_6$ band at 0.5\,eV below the Fermi level, has the spin up component at lower energies indicating the ferromagnetic sign of the exchange interaction with Mn spins. Instead, the carrier spins in the $\Gamma_8$ bands are antiferromagnetically coupled to Mn spins.

\subsection{Spin splitting along the \textbf{k} path
 without spin-orbit coupling}

\begin{figure}[tb]
\includegraphics[width=9.4cm,angle=0]{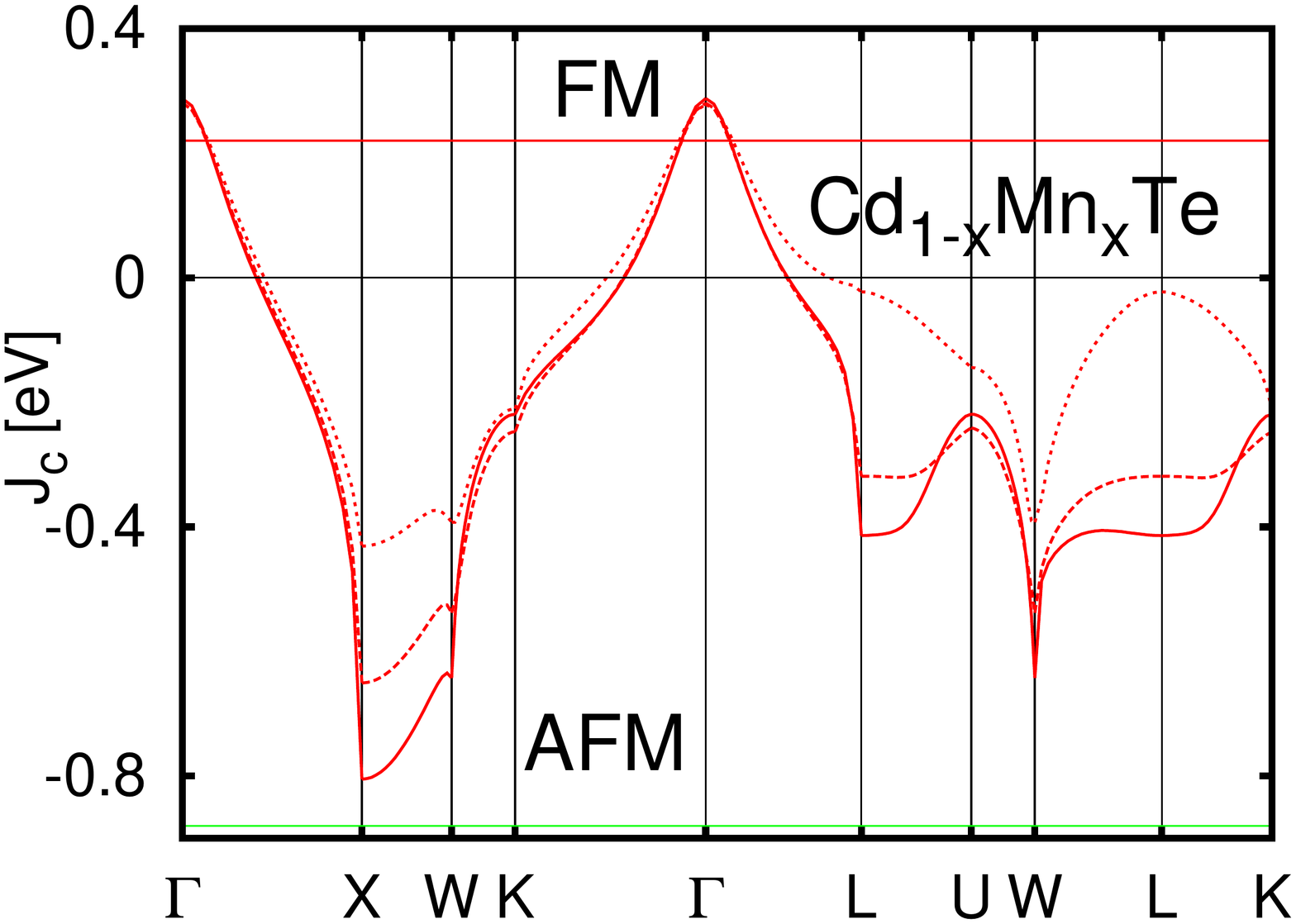}
\includegraphics[width=9.4cm,angle=0]{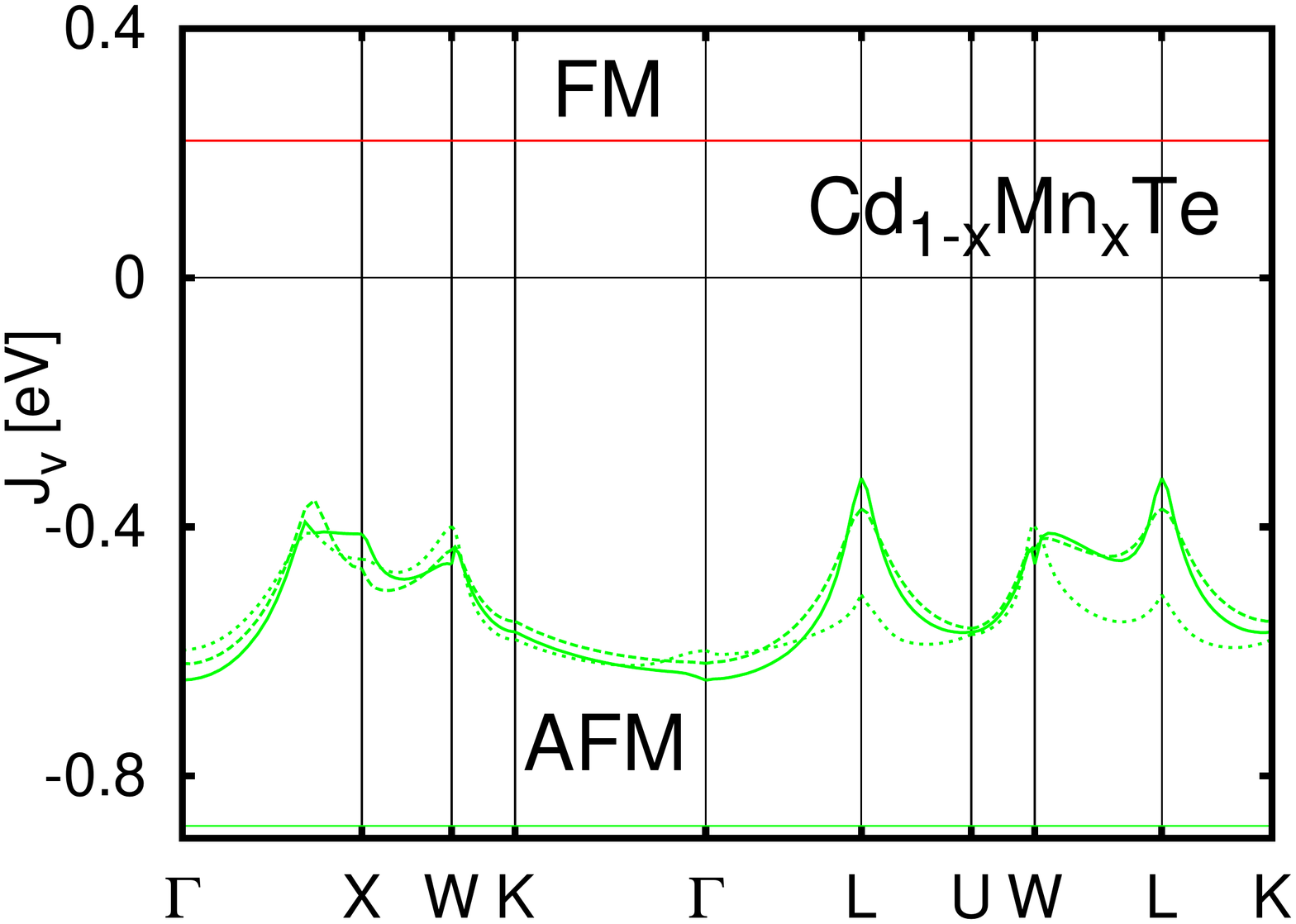}
\caption{GGA values of effective exchange integrals for the conduction and valence band,  $J_c$ in the top panel (red lines) and $J_v$ in the bottom panel (green lines) for Cd$_{1-x}$Mn$_{x}$Te compared to experimental values at the $\Gamma$ point, determined by exciton magnetospectroscopy, represented by horizontal lines \cite{Gaj:1979_SSC}. The experimental effective exchange integral is positive for the conduction band and negative for the valence at the $\Gamma$ point.
The solid, dashed, and dotted lines represent the band spin splitting for entirely spin-polarized Mn ions with concentrations $x =2/64 =0.03125, 4/64 =0.0625$ and $8/64 =0.125$, respectively. The computations have been performed neglecting spin-orbit coupling and for $U_{\text{Mn}} =5$\,eV.}
\label{fig:SPINPLITNOSOC_a324}
\end{figure}

To take random Mn positions into account, we have used the SQS for determining the ${\bm k}$-dependence of exchange energies $J_c$ and $J_v$. Figure \ref{fig:SPINPLITNOSOC_a324} shows $J_c(\bm{k})$ and $J_v(\bm{k})$ computed for  Cd$_{1-x}$Mn$_{x}$Te with various Mn concentrations $x$ and $U_{\text{Mn}}=5$\,eV. In agreement with the experimental results \cite{Gaj:1979_SSC}, the values determined for the $\Gamma$ point do not depend on $x$, and their DFT values, $N_0\alpha = 0.28$\,eV and $N_0\beta = -0.63$\,eV,  describe the sign, and also reasonably well the experimental magnitudes, $N_0\alpha = 0.22$\,eV and $N_0\beta = -0.88$\,eV \cite{Gaj:1979_SSC}, depicted by horizontal lines in Fig.~\ref{fig:SPINPLITNOSOC_a324}. The exchange splittings at $\Gamma$ means that there is a large energy difference between transitions from the two heavy hole subbands (or for the creation of the heavy hole excitons). This giant Zeeman splitting is described by $\Delta E = xS(N_0\alpha- N_0\beta)$.

As seen, $J_v$ remains negative (antiferromagnetic) for all $\bm{k}$-values, and its magnitude slightly oscillates along the $\bm{k}$-path, it reaches a maximum at $\Gamma$ and a minimum at $L$ for small Mn concentrations,
and between the $\Gamma$ and $X$ points at the highest studied $x=8/64=0.125$.

In contrast to $J_v$, $J_c$ changes sign and is highly oscillating along the $\bm{k}$-path:
the sign of $J_c$ is positive (ferromagnetic potential $s$-$d$ exchange) at the $\Gamma$ point, in which the conduction band wave function has the $s$-type character, but becomes antiferromagnetic away from the $\Gamma$ point. This behavior originates from an admixture of anion wave functions to the Bloch amplitudes $u_{\bm{k}}$ and, thus, from a significant role of antiferromagnetic $sp$-$d$ kinetic exchange, affected by the phase factors $\exp(i\bm{k}\cdot\bm{r})$, known from the Kondo physics in dilute magnetic metals \cite{Schrieffer:1966_PR}.
In the negative sign region, the absolute value of $J_c$ reaches a maximum at the $X$ points and a minimum at the $U$ points at small concentrations and at the $L$ points for $x=0.125$. Such a dependence results from an increase of $sp$-$d$ hybridization and, thus, of the kinetic exchange if a given state approaches the $3d$ Mn shell, in the $J_c$ case, the upper Hubbard $3d^6$ band. In agreement with this interpretation,  exchange energies at the boundary of the Brillouin zone  get reduced when we increase $U_{\text{Mn}}$ because the Mn $d$-states move away from the relevant bands, and the hybridization between the host bands and the $3d$ shells of Mn-impurities becomes suppressed.

We are interested in the origin of a large reduction in magnitudes of exchange splittings at $L$ compared to $\Gamma$, as determined by interband magnetooptical studies. The corresponding reduction factor can then be defined as
\begin{equation}
r = \frac{J_c(\Gamma)-J_v(\Gamma)}{J_c(L)-J_v(L)}.
\end{equation}
As seen, the sign of $r$ is determined by relative magnitudes of $J_c$ and J$v$, whose values and signs at the Brillouin zone boundary depend on a subtle competition between positive (ferromagnetic) potential and negative (antiferromagnetic) kinetic exchange energies.  According to  {\em ab initio} results presented in Fig.~\ref{fig:SPINPLITNOSOC_a324},  $J_c$ and $J_v$ at the $L$ points have the same sign and similar magnitudes. This fact explains qualitatively why the experimentally observed splitting of optical spectra is relatively small at $L$  compared to $\Gamma$ \cite{Ginter:1983_SSC,Coquillat:1986_SSC,Ando:2011_JAP}. Given our results, the previous attempt to interpret the large value of $r$  was quantitatively unsuccessful because a strong dependence of $J_c$ on $\bm{k}$ was disregarded \cite{Bhattacharjee:1990_PRB}. At the same time, our data suggest a relatively strong dependence of $J_c$ and $J_v$ at $L$ on $x$. Experimental results accumulated so far do not  corroborate this expectation.

Finally, we mention the relevance of our {\em ab initio} results for magnetooptical studies of (001)Cd$_{1-x}$Mn$_{x}$Te quantum wells sandwiched between  Cd$_{1-x-y}$Mn$_{x}$Mg$_{y}$Te barriers \cite{Mackh:1996_PRB,Merkulov:1999_PRL}, interpreted theoretically by a $k \cdot p$ model \cite{Merkulov:1999_PRL}. Experimental data pointed to a reduction of exchange splittings compared to those found for bulk samples, the behavior assigned to effectively non-zero values of $k\sim \pi/d$ ($d$ is the quantum well thickness) at which splittings were probed  \cite{Mackh:1996_PRB,Merkulov:1999_PRL}.  Our evaluation, making use of data in Figs.~\ref{fig:SPINPLITNOSOC_a324} and \ref{fig:CdMnTe_bands} for the relevant ${\bm k}$-direction (the $\Gamma$-$X$ line) indicates that the decrease of $J_c$ with $\bm{k}$ is consistent with the  experimentally observed and theoretically described decrease of $N_0\alpha$ with diminishing $d$, if penetration of the wave function into barriers is taken into account \cite{Merkulov:1999_PRL}.

\begin{figure}[tb]
\includegraphics[width=9.4cm,angle=0]{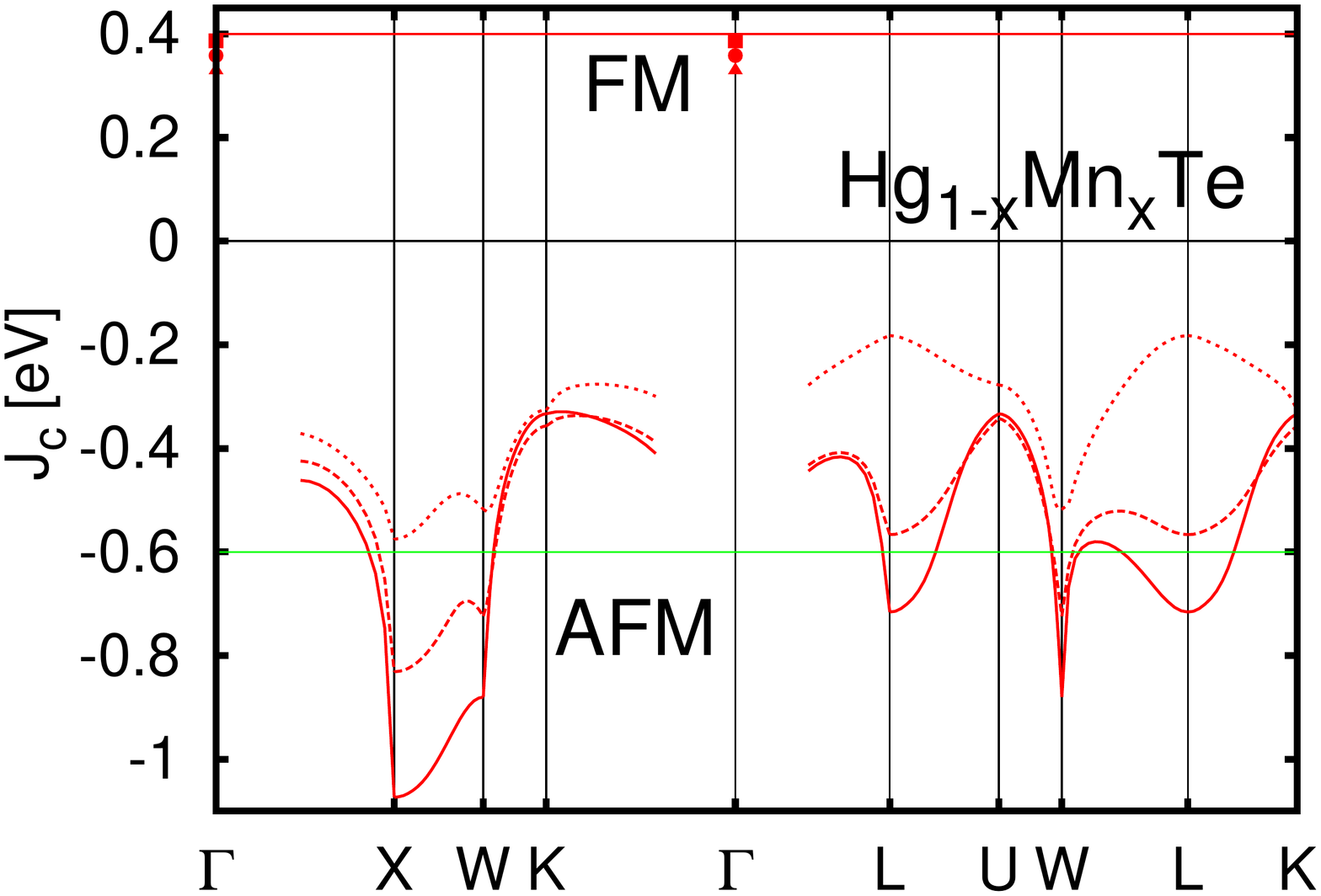}
\includegraphics[width=9.4cm,angle=0]{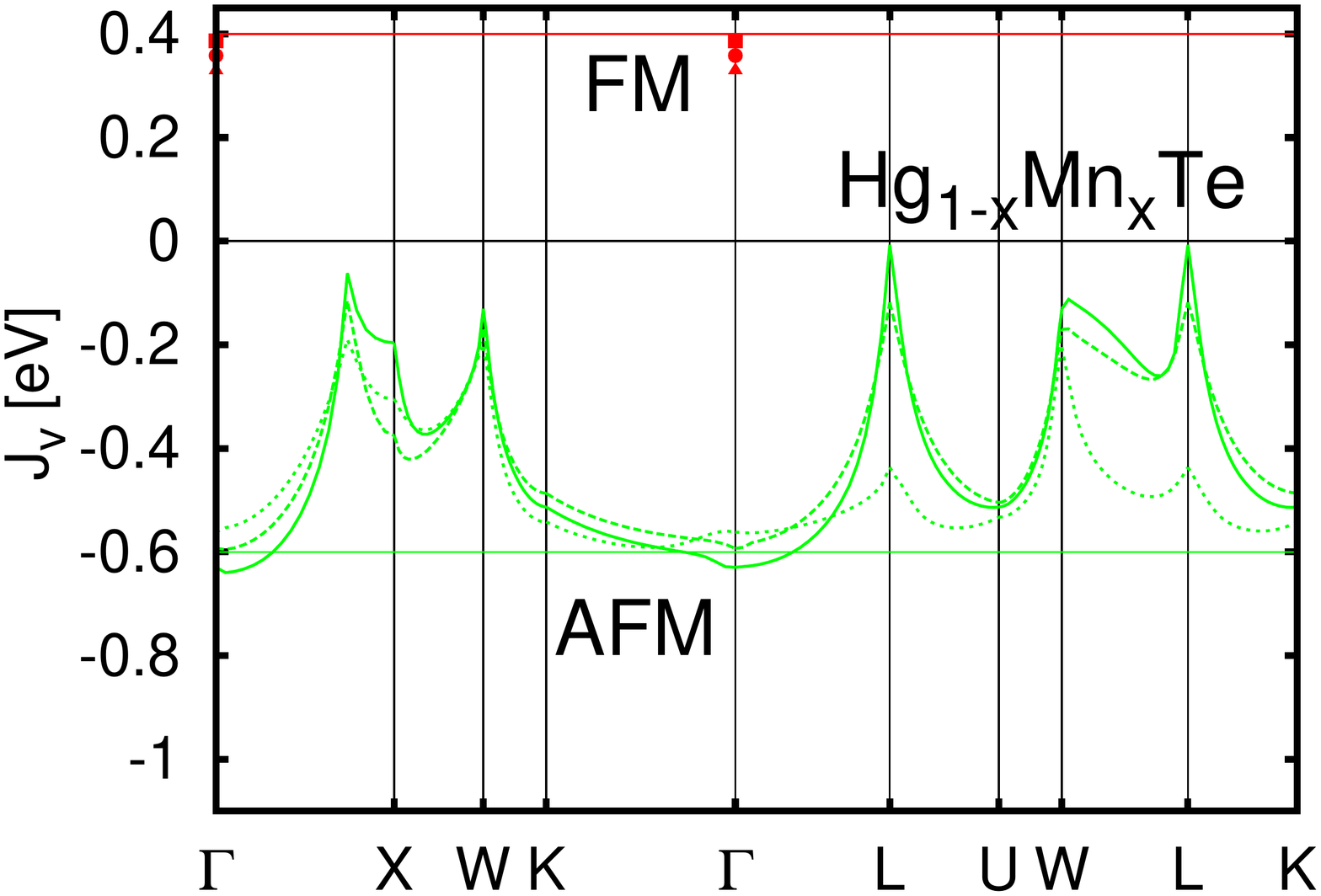}
\caption{The values of effective exchange integrals for the  conduction and valence band,  $J_c$ in the top panel (red lines) and $J_v$ in the bottom panel (green lines) for Hg$_{1-x}$Mn$_{x}$Te compared to experimental values at the $\Gamma$ point represented by horizontal lines \cite{Furdyna:1988_JAP,Dobrowolska:1981_JPC}.
The sign of the experimental effective exchange integral is negative for the valence band and positive for the conduction band at the $\Gamma$ point.
The solid, dashed, and short-dashed represent the spin splitting for $x=2/64 =0.03125, 4/64 =0.0625$ and $8/64 =0.125$, respectively.
The square, circle and triangle represent an effective exchange integral of the $\Gamma_6$ band for $x=2/64 =0.03125, 4/64 =0.0625$, and $8/64 =0.125$, respectively.}
\label{fig:SPINPLITNOSOC_Hg}
\end{figure}

Figure~\ref{fig:SPINPLITNOSOC_Hg} shows $J_c(\bm{k})$ and $J_v(\bm{k})$ extracted from the band structure computations without SOC for Hg$_{1-x}$Mn$_x$Te with different values of $x$. In the vicinity of the zone center we present single data points corresponding to the exchange energy of the $\Gamma_6$ band, i.e., $N_0\alpha$. The trends in $\bm{k}$ dependencies are similar to the Mn-doped CdTe. In particular, $J_v$ stays negative in the whole Brillouin zone and $J_c(\bm{k})$ becomes negative away from the zone center.

On the experimental side, there are two sets of the determined $N_0\alpha$ and $N_0\beta$ values, differing by more than a factor of two, in the case of Hg$_{1-x}$Mn$_x$Te \cite{Dietl:1994_B}. Our computational results point to the lower values, i.e., $N_0\alpha = 0.4$\,eV and $N_0\beta =-0.6$\,eV \cite{Furdyna:1988_JAP,Dobrowolska:1981_JPC}. Furthermore, according to experimental findings,
magnetic circular dichroism at $L$  has the same sign for  Hg$_{1-x}$Mn$_x$Te as found for Cd$_{1-x}$Mn$_x$Te at $L$ and at $\Gamma$, independently of Mn content $x$ \cite{Coquillat:1989_PRB}. Our data suggest the opposite sign since, according  to the results in Fig.~\ref{fig:SPINPLITNOSOC_Hg}, $J_c(L)-J_v(L) <0$ for Hg$_{1-x}$Mn$_x$Te in the relevant effective Mn concentrations, $x \lesssim 6$\%.

In summary, the DFT results presented in Figs.~\ref{fig:SPINPLITNOSOC_a324} and \ref{fig:SPINPLITNOSOC_Hg}, obtained without taking SOC into account, have qualitatively shown how exchange spin-splitting of bands evolves with the ${\bm{k}}$-vector spanning the whole Brillouin zone. This dependence reflects (i) the $\bm{k}$-dependent mixing between cation and anion wave functions,  which affects a relative contribution of the potential and kinetic components to the $sp$-$d$ exchange and (ii) the energy position of a given ${\bm{k}}$ state with respect to the open Mn $d$ shells, which controls the magnitude of the $\bm{k}$-dependent kinetic exchange.  Quantitatively, however, bands' energy position and, thus, the magnitude of exchange splitting depends significantly on SOC. Moreover, in the presence of SOC, exchange splitting of a given band state changes with the orientation of its ${\bm{k}}$-vector with respect to the direction of ${\bm{M}}(T,{\bm{H}})$. This means that, in general, exchange splitting of particular $L$ valleys differs, depending on the angle between ${\bm{k}}_L$ and ${\bm{M}}(T,{\bm{H}})$.  Furthermore, under non-zero magnetization ${\bm{M}}(T,{\bm{H}})$,  degeneracy of states with different projections of the orbital momentum is removed in the presence of SOC.  This results in MCD, i.e., different transition probabilities for two circular light polarizations $\sigma^+$ and $\sigma^-$. In other words, MCD vanishes in the absence of SOC and, neglecting the magnetic field's direct effect on electronic states, in the absence of exchange-induced spin splittings.

We are interested in interpreting MCD  for Cd$_{1-x}$Mn$_x$Te and Hg$_{1-x}$Mn$_x$Te, taken at photon energies corresponding  to free excitons at the fundamental gap at the $\Gamma$ point ($E_0$ and $E_0 + \Delta_0$ transitions) and at the $L$ points ($E_1$ and $E_1 + \Delta_1$ transitions) \cite{Ginter:1983_SSC,Coquillat:1986_SSC,Ando:2011_JAP,Coquillat:1989_PRB}, where $\Delta_0$ and $\Delta_1$ are the spin-orbit splitting of the valence band at the $\Gamma$ and $L$ points of the Brillouin zone, respectively. Our theoretical approach considering SOC involves several steps. First, we use the DFT calculations with SOC taken into account to determine the parameters of a tight-binding model for CdTe and HgTe. Second, we consider the Mn-doped case and obtained from DFT on-site and hopping energies for Mn $d$ shell and its coupling to band states in CdTe and HgTe. Third, these parameters are  incorporated into $sp$-$d$ exchange Hamiltonian that takes into account the presence of $\bm{k}$-dependent kinetic and potential exchange interactions in the molecular-field and virtual-crystal approximations suitable for Cd$_{1-x}$Mn$_x$Te and Hg$_{1-x}$Mn$_x$Te.
In the fourth step, we use this model to determine energies  of optical transitions at the $\Gamma$ and $L$ points of the Brillouin zone. We then develop the $\bm{k} \cdot \bm{p}$ theory for the $L$ point of the Brillouin zone in zinc-blende semiconductors, which allows as determining the oscillator strength  for particular transitions and two circular light polarizations. With this information, we are in a position to compute the  MCD spectra for Cd$_{1-x}$Mn$_x$Te and Hg$_{1-x}$Mn$_x$Te, and validate our approach by comparison to experimental data.

\subsection{DFT with  spin-orbit coupling and minimal tight-binding model for CdTe and HgTe}

\begin{figure*}[tb]
\includegraphics[width=8.9cm,angle=0]{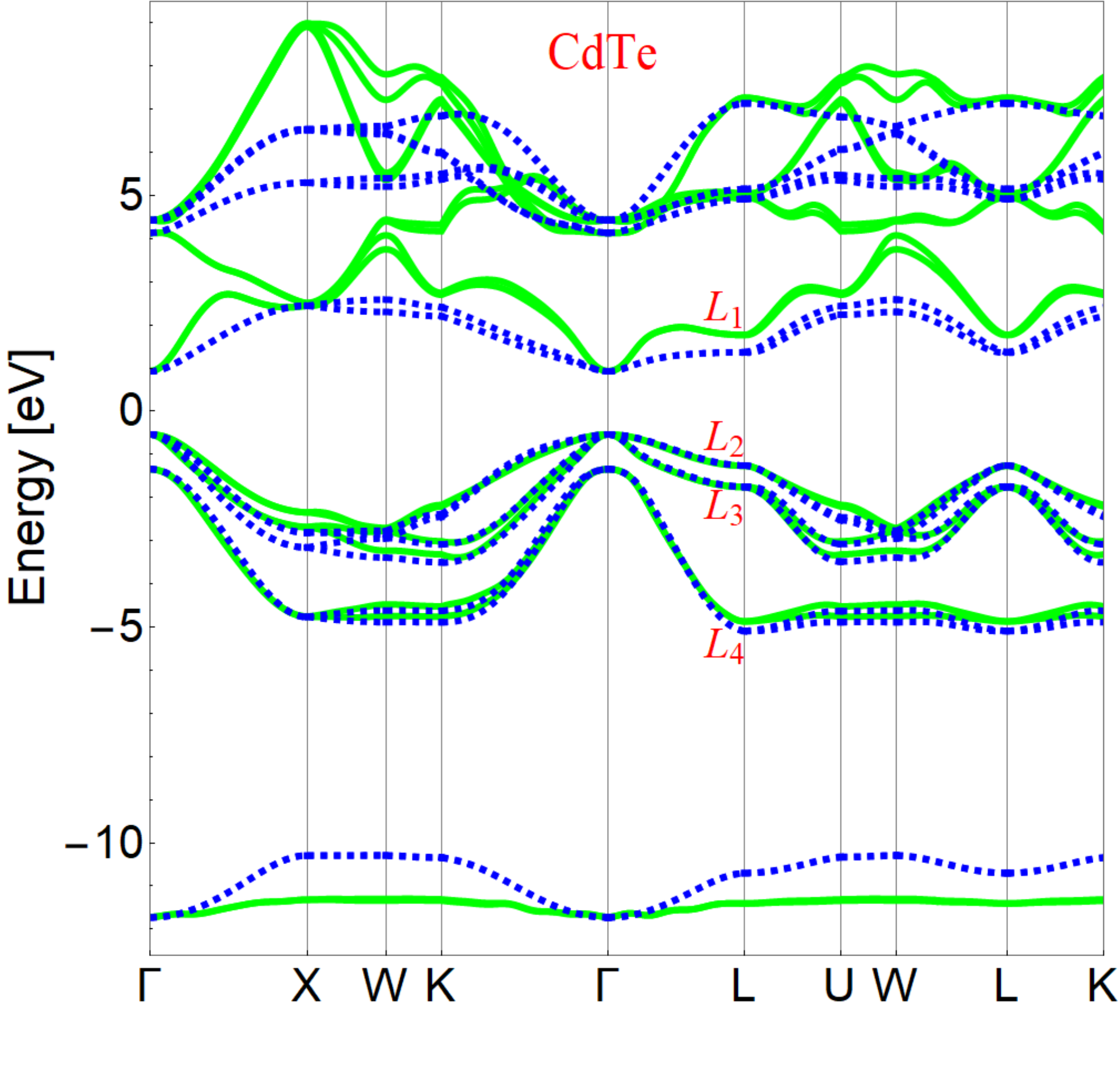}
\includegraphics[width=8.9cm,angle=0]{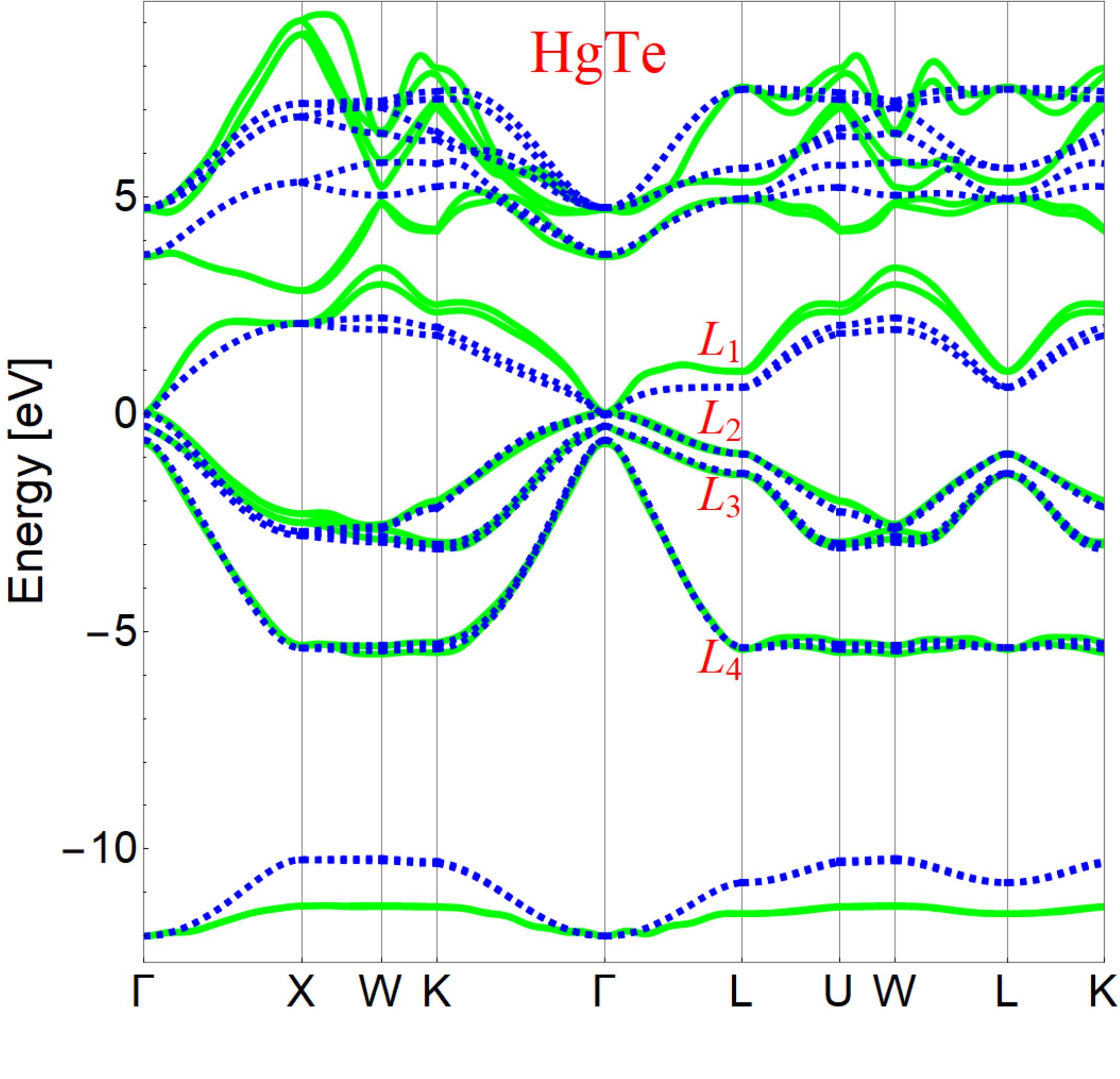}
\caption{Comparison between the Wannier bands obtained by MBJLDA (solid green line) and our tight-binding model (dashed blue lines) for CdTe (left panel) and HgTe (right panel) taking spin-orbit coupling into account. The bands labeled $L_1$,$L_2$,$L_3$, and $L_4$ are described by irreducible representations of the double group $L_6$, $L_{4,5}$,
$L_6$, and $L_6$, respectively. }
	\label{fig:CdTeHgTe}
\end{figure*}

 As mentioned in Sec.~\ref{sec:methodology}, we use  MBJLDA to determine the relativistic band structure of CdTe and HgTe
with experimental lattice constants 6.48\,{\AA} for CdTe and 6.46\,{\AA} for HgTe.
To extract energy dispersions $E(\bm{k})$ of the electronic bands, the Slater-Koster interpolation scheme is employed. The obtained results are shown in Fig.~\ref{fig:CdTeHgTe}. The computed values of energy gaps and spin-orbit splittings for CdTe and HgTe are summarized in Table~\ref{tab:CdTeHgTe}, and show good agreement with experimental data.

We aim to use the {\em ab initio} results for determining the parameters of the one-electron Hamiltonian in the tight-binding approximation (TBA), which will adequately describe the band structure and $sp$-$d$ exchange splittings of bands at an arbitrary $\bm{k}$-point of the Brillouin zone with SOC taken into account. Similar to the previous descriptions of CdTe and HgTe within TBA \cite{Tarasenko:2015_PRB}, we consider $sp^3$ orbitals per atom and the nearest neighbor hopping. In particular, from positions of the electronic bands at $\Gamma$  we obtain the TBA on-site energies and the spin-orbit splittings. Then we use as constraints the DFT values of the band energies at the $\Gamma$, $X$, and $L$ points. We create an equation system and search for the values of the hopping energies $V$. If this procedure results in multiple solutions, we select $V_{sp}$ that has the same sign as the first-neighbor hopping energy among Wannier functions.  The TBA parameters obtained in this way are shown in Table \ref{tab:tab1}. Since the atomic radius of Cd is smaller than of Hg, whereas the bond length is greater in CdTe compared to
HgTe, there are no systematic differences in the magnitudes of the hybridizations $V$ between these two compounds.
Figure~\ref{fig:CdTeHgTe} presents a comparison of the band structures resulting from the MBJLDA and our TB model.

\begin{table}
\caption{Energy gaps and spin-orbit splittings (in eV) at the $\Gamma$ and $L$ points,
where $E_0 = E(\Gamma_6)-E(\Gamma_8)$ and $\Delta_0 = E(\Gamma_8)- E(\Gamma_7)$
and
at the $L$ points of the Brillouin zone,
where $E_1 = E(L_{1})-E(L_{2})$ and $\Delta_1 = E(L_{2})-E(L_{3})$ for CdTe and HgTe, as determine from MBJLDA (our data), ETB \cite{Tarasenko:2015_PRB}, our TB model,
and experimentally. Labeling of $L$ points is shown in Fig.~\ref{fig:CdTeHgTe}.}
\vspace{0.2cm}
\begin{tabular}{c|c|c|c|c}
\hline \hline
              \multicolumn{5}{c}{CdTe}   \\  \hline
                 & MBJLDA & ETB     & TB     & expl   \\  \hline
   $E_{0}$       &1.47&  1.56   & 1.46  &   1.61   (Ref.\,\onlinecite{Laurenti:1990_JAP})\\
   $\Delta_{0}$  &0.81&  0.78   & 0.80  &   0.95 ({Ref.\,\onlinecite{Twardowski:1980_SSC}}) \\
   $E_{1}$       &3.03&  4.78   & 2.63  &   3.28 (Ref.\,\onlinecite{Cardona:1967_PR}); 3.46 (Ref.\,\onlinecite{Chadi:1972_PRB})   \\
   $\Delta_{1}$  & 0.49 &  0.47   & 0.49     & $0.6\pm 0.05$ (Ref.\,\onlinecite{Cardona:1967_PR}) \\
   \hline
 \end{tabular}
\vspace{0.2cm}

\begin{tabular}{c|c|c|c|c}
\hline
              \multicolumn{5}{c}{HgTe}   \\  \hline
               & MBJLDA & ETB     & TB     & expl   \\  \hline
 $E_{0}$       &  -0.31 & -0.22   &  -0.27  & -0.30 (Ref.\,\onlinecite{Laurenti:1990_JAP}) \\
  $\Delta_{0}$ &{ }0.70 &{ }0.72  &{ }0.59  &{ }0.91 (Ref.\,\onlinecite{Sakuma:2011_PRB})\\
  $E_{1}$      &{ }1.89 &{ }2.75  &{ }1.53  &{ }2.12 (Ref.\,\onlinecite{Ando:2011_JAP}); 2.25 (Ref.\,\onlinecite{Chadi:1972_PRB}) \\
 $\Delta_{1}$  &{ }0.49 &{ }0.50  &{ }0.45  &{ }0.62--0.75 (Ref.\,\onlinecite{Sakuma:2011_PRB})\\
 \hline
\end{tabular}
 \label{tab:CdTeHgTe}
\end{table}

\begin{table}
\caption{Values of the on-site $E$, hopping $V$, and spin-orbit splitting $\Delta$ energies (in eV) of our minimal tight-binding model for CdTe and HgTe, which includes $sp^3$ orbitals of anions $a$ and cations $c$, and the nearest-neighbor hopping.  Zero energy is set at the top of the valence band.}
\vspace{0.2cm}
\begin{tabular}{ c|c| c }
\hline\hline
                       & CdTe & HgTe  \\ \hline
   $E_{s(a)}$            &   $-8.7752$             &   $-9.1555$  \\ 
   $E_{s(c)}$            &   $-0.9526$             &   $-3.1156$  \\ 
   $E_{p(a)}$            &   $-0.2669$             &   $-0.1742$  \\ 
   $E_{p(c)}$            & $\hspace{0.3cm}4.8663$  & $\hspace{0.3cm}4.3691$ \\ 
   $V_{ss}\sigma$        &   $-1.2431$             &   $-1.2569$  \\ 
   $V_{s(a)p(c)}\sigma$  & $\hspace{0.3cm}1.6379$  &  $\hspace{0.3cm}1.7229$ \\ 
   $V_{s(c)p(a)}\sigma$  & $\hspace{0.3cm}1.5463$  &  $\hspace{0.3cm}1.4834$  \\ 
   $V_{pp}\sigma$        & $\hspace{0.3cm}2.0139$  &  $\hspace{0.3cm}2.2132$ \\ 
   $V_{pp}\pi$           & $-0.9875$              &   $-0.9830$ \\ 
   $\Delta_{a}$          & $\hspace{0.3cm}0.8025$  & $\hspace{0.3cm}0.5915$ \\ 
   $\Delta_{c}$          & $\hspace{0.3cm}0.2925$  & $\hspace{0.3cm}1.0824$ \\ \hline
\end{tabular}
\label{tab:tab1}
\end{table}

By construction, the TB parameters collected in Table~\ref{tab:tab1} lead to similar bandgaps and spin-orbit splittings as determine by DFT, and displayed in Table~\ref{tab:CdTeHgTe} and Fig.~\ref{fig:CdTeHgTe}. These parameters describe well experimental energy gaps at both $\Gamma$ and $L$. For comparison, we present in Table~\ref{tab:CdTeHgTe} the magnitudes of bandgaps and spin-orbit splittings computed by using the tight-binding parameters determined by Tarasenko {\em et al.} \cite{Tarasenko:2015_PRB} in reference to experimental data primarily at $\Gamma$.

\subsection{Tight-binding parameters from DFT for Cd$_{1-x}$Mn$_x$Te and Hg$_{1-x}$Mn$_x$Te}
\label{sec:TBA_Mn}

We are interested in evaluating Slater-Koster parameters associated with the presence of open
 $3d$ shells of Mn in Cd$_{1-x}$Mn$_x$Te and Hg$_{1-x}$Mn$_x$Te, i.e.,
 hopping energies between Mn $3d$ orbitals and $5sp^3$ states of the nearest neighbor Te anions
as well as energetic positions of Mn $d$ levels. For this purpose, we use supercells
with $2 \times 2 \times 2$ unit cells, each containing one Mn atom. The  GGA+U technique is employed with  $U_{\text{Mn}}=5$\,eV
and J$_{\text{Hund}}=0.75$\,eV as well as with the PBE exchange functional. Since in such alloys, no $E(\bm{k})$ dependencies
can be derived, we extract the
Slater-Koster parameters $V$ directly from the hopping energies among the relevant Wannier functions, which means that their accuracy
is presumably of the order of 20\%. The  magnitudes of determined parameters
are shown in Table \ref{tabMnCdTe2}. A lower position (by about 0.3\,eV) of $d$ levels in HgTe compared to CdTe originates
from the valence band offset between these two compounds \cite{Kowalczyk:1986_PRL,Dietl:1988_PRB}. The spin-up Mn states are more localized
and the hopping energies $V$ related to $d\uparrow$ are smaller. On the other hand,  noticeable
dissimilarities in hopping energies of the two compounds are caused by differences in the bond length and in the participation of
cation orbitals to the $s$-like and $p$-like wave functions.

\begin{table}[tb]
\caption{The DFT values of on-site and hopping energies (in eV) for Mn $3d$ orbitals ($t_{2g}$ and $e_g$) and the nearest-neighbor Te $5s$ and $5p$ states for Cd$_{1-x}$Mn$_x$Te and Hg$_{1-x}$Mn$_x$Te. Zero energy is set at the top of the valence band.}
\vspace{0.2cm}
\begin{tabular}{c |c| c} \hline\hline
 & Cd$_{1-x}$Mn$_x$Te & Hg$_{1-x}$Mn$_x$Te \\ \hline
   $E_{t2g\uparrow}$         &  $-4.702$             &  $-4.997$  \\
   $E_{t2g\downarrow}$       &$\hspace{0.3cm}2.198$  &$\hspace{0.3cm}1.821$   \\
   $E_{eg\uparrow}$          &  $-4.525$             &  $-4.858$   \\
   $E_{eg\downarrow}$        &$\hspace{0.3cm}2.665$  &$\hspace{0.3cm}2.293$  \\
   $V_{sd}\uparrow\sigma$    &  $-1.081$             &  $-1.232$  \\
   $V_{sd}\downarrow\sigma$  &  $-1.949$             &  $-1.957$  \\
   $V_{pd}\uparrow\sigma$    &  $-0.488$             &  $-0.364$  \\
   $V_{pd}\downarrow\sigma$  &  $-0.957$             &  $-0.987$  \\
   $V_{pd}\uparrow\pi$       &$\hspace{0.3cm}0.253$  &$\hspace{0.3cm}0.187$  \\
   $V_{pd}\downarrow\pi$     &$\hspace{0.3cm}0.844$  &$\hspace{0.3cm}0.854$  \\ \hline
\end{tabular}
 \label{tabMnCdTe2}
\end{table}

\subsection{Tight-binding model with $sp$-$d$ exchange interaction}

We now present and discuss the tight-binding Hamiltonian with four $sp^3$ orbital per atom containing a term describing giant Zeeman splitting of bands in the presence of spin polarized Mn ions. This splitting is brought about by: (i) the kinetic exchange resulting from spin-independent hybridization between Mn $3d$ shells and band states derived from the $5s$ and $5p$ orbitals of the four neighboring Te anions; (ii) direct (potential) exchange coupling  of electrons residing on the open Mn $3d$ shell to band carriers visiting Mn $4s$ or $4p$ orbitals. Our approach is developed within the molecular-field and virtual-crystal approximations, and generalizes previous descriptions of DMSs within TBA \cite{Oszwaldowski:2006_PRB,Sankowski:2007_PRB} by taking into account the $\bm{k}$-dependence of the kinetic exchange according to,
\begin{equation}
  {\cal{H}}({\bm{k}}) = {\cal{H}}_{\text{TB}}({\bm{k}}) + {\cal{H}}_{sp-d}({\bm{k}}).
\label{eq:H_TBA}
\end{equation}
Within our model ${\cal{H}}_{\text{TB}}(\bm{k})$ is a $16 \times 16$ matrix, with on-site energies of $s$ and $p$ cation and anion orbitals on the diagonal; $\bm{k}$-dependent hopping $t_{mn}(\bm{k})$ between orbitals $m$ and $n$ of the nearest-neighbor (n.n.) cation-anion pairs, and the intraatomic spin-orbit term involving $p$-type orbitals of the cation and anion, respectively,
\begin{equation}
  \left< m, s \middle| {\cal{H}}_{\text{TB}}(\bm{k}) \middle| n, s' \right> = E_n \delta_{mn} + t_{mn}(\bm{k})+ \frac{\Delta_{a(c)}}{3} \sum_{\alpha} I'^{\alpha}_{mn} \sigma^{\alpha}_{ss'},
  \label{eq: tkba}
\end{equation}
where $t_{mn}(\bm{k})$ is the total hopping energy to the four n.n. atoms in the zinc-blende lattice including $\bm{k}$-dependent phase factors,
\begin{equation}
  t_{mn}(\bm{k}) = \sum_{\bm{R}_m \in \text{n.n.}(\bm{R}_n)} V_{mn}(\bm{R}_m - \bm{R}_n) \exp[i \bm{k} \cdot (\bm{R}_m - \bm{R}_n)]. \label{eq: tk}
\end{equation}
The Slater-Koster interatomic matrix elements (dependent on the direction cosines of the vector from the location $\bm{R}_n$ of the orbital $n$
to the location $\bm{R}_m$ of the orbital $m$) are denoted as $V_{mn}(\bm{R}_m - \bm{R}_n)$, and their values
for various combinations of orbitals ($ss\sigma$, $sp\sigma$, $pp\sigma$, $pp\pi$) are given in Table~\ref{tab:tab1}.
The intraatomic spin-orbit splitting energies of the anion (cation) $p$ states are denoted by $\Delta_{a(c)}$, respectively,
the orbital momentum operator $I'^{\alpha}_{\beta\gamma}$ in the Cartesian basis  ($\alpha, \beta, \gamma = x, y, z$)
can be written using the Levi-Civita symbol $\epsilon_{\alpha\beta\gamma}$ as
$I'^{\alpha}_{\beta\gamma} = -i \epsilon_{\alpha\beta\gamma}$, and $(\sigma^{\alpha})_{\alpha = x, y, z}$ stand for the set of Pauli matrices.

The exchange interaction is taken into account in the molecular-field and virtual-crystal approximations: the weight, by which Mn spin polarization affects the band splittings, is described by the vector $\bm{X} = x_{\text{eff}} \bm{S}$, where $x_{\text{eff}}=x$ and $S = 5/2$ if all Mn ions are spin polarized, and the direction of $\bm{S}$ is imposed by the external magnetic field. Hence, the vector $\bm{X}$ is related to Mn spin magnetization according to $\bm{M} = -g_{\text{Mn}}N_0\bm{X}$, where $g_{\text{Mn}} = 2.0$ and $N_0$ is the cation concentration. Then, the relevant $sp$-$d$ exchange Hamiltonian, to be added to the TB Hamiltonian, assumes the form:
\begin{widetext}
\begin{equation}
  \left< m, s \middle| {\cal{H}}_{sp-d}(\bm{k}) \middle| n, s' \right> = -\frac{1}{2} \sum_{\alpha} X^{\alpha} \sigma^{\alpha}_{ss'} \left[ \frac{1}{S}\sum_{d} \left(\frac{1}{E_{d\uparrow} - E_{\bm{k}}} -\frac{1}{E_{d\downarrow}- E_{\bm{k}}}\right)t_{md}(\bm{k}) t_{dn}(\bm{k}) + \left< m \middle|J_{4s-3d} \hat{P}_{sc} + J_{4p-3d} \hat{P}_{pc} \middle| n \right> \right]
  \label{eq:sp-d}
\end{equation}
\end{widetext}
The first term in the brackets was given by Schrieffer and Wolff \cite{Schrieffer:1966_PR}, and accounts for the kinetic exchange; this contribution
is $\bm{k}$-dependent (via $E_{\bm{k}}$ and $t_{md}(\bm{k})t_{dn}(\bm{k})$), and may be non-diagonal. In this term, the $d$ index runs over the $t_{2g}$ and $e_g$ orbitals of Mn, the matrix of hoppings is defined as in Eg.\,(\ref{eq: tk}), but this time takes into account the matrix of hoppings between cation $d$ orbitals and n.n. anion $s$ and $p$ orbitals, as given by Schrieffer and Wolff's canonical transformation that is equivalent to the second order perturbation theory. Accordingly, as input parameters one should adopt the values unperturbed by the $sp$-$d$ hybridization. To this end, as $E_{d\uparrow,\downarrow}$ we take $E_{eg\uparrow,\downarrow}$, as $e_g$ states hybridize weakly in the tetrahedral case \cite{Wei:1987_PRB}, and as $V_{nd}$ spin average values of $V_{pd}\sigma$ and $V_{pd}\pi$ in Table~\ref{tabMnCdTe2}. If relevant $d$ and ${\bm k}$ states cross, higher order perturbation theory is necessary \cite{Sliwa:2018_PRB}.

The second term in Eq.\,(\ref{eq:sp-d}) describes intra-Mn direct (potential) exchange $J_{4s-3d}$ and $J_{4p-3d}$ between electrons residing on $3d$ and $4s$ or $4p$ Mn states, respectively; $\hat{P}_{sc}$ and $\hat{P}_{pc}$ are the projectors on cation $s$ and $p$ states. This ferromagnetic potential exchange assumes the canonical Heisenberg form, $-J\bm{s} \cdot \bm{S}$.
According to spectroscopic studies, $J_{4s-3d} = 0.392$\,eV and $J_{4p-3d} = 0.196$\,eV for free Mn$^{+1}$  ions \cite{Dietl:1994_PRB}. The values of potential exchange are reduced in compound semiconductors by admixtures of anion orbitals to the Bloch wave functions taken into account within TBA and, possibly, also by screening (neglected here).

By incorporating Eq.\,(\ref{eq:sp-d}) into the TB Hamiltonian we obtain ${\bm{k}}$-dependent splittings of bands for a given direction of Mn magnetization ${\bm{M}}$,
from which the magnitudes of $sp$-$d$ exchange energies $J$, i.e.,
band splittings divided by $Sx_{\text{eff}}$ can be determined, as defined in Eq.\,(\ref{eq: J_cJ_v}).
Table \ref{tab:J_cJ_v} shows the magnitudes of $J$  for conduction and valence bands at the $\Gamma$ and $L$ points of the Brillouin zone, relevant to interband magnetooptical transitions $E_0$, $E_0+\Delta_0$, $E_1$, and $E_1+ \Delta_1$, respectively.
These values have been obtained for $x_{\text{eff}} = 0.0625$ and making use of the TB parameters determined from DFT and collected in Tables \ref{tab:tab1} and \ref{tabMnCdTe2}. Particular $J$ values have been determined by independent diagonalization of the TB Hamiltonian containing $E_{\bm{k}}$ in the kinetic exchange term (Eq.\,\ref{eq:sp-d}) corresponding to the band extremum in question, computed by diagonalizing the TB Hamiltonian without $sp$-$d$ exchange.

\begin{table}
\caption{Exchange energies $J(\bm{k})$ (in eV) for the conduction and valence bands
at the $\Gamma$ and $L$ points of the Brillouin zone computed for Cd$_{1-x}$Mn$_x$Te and Hg$_{1-x}$Mn$_x$Te
with $x_{\text{eff}} = 0.0625$ employing  tight-binding parameters displayed in Tables \ref{tab:tab1} and \ref{tabMnCdTe2}.
In the DMS nomenclature, $J(\Gamma_6)$ and $J(\Gamma_8)$ correspond to $N_0\alpha$ and $N_0\beta$, respectively.
Since for the $L_2$ and $L_3$ case the values of $J$ are anisotropic, we show in this Table the data are for the magnetization vector ${\bm{M}} \| [111]$
and for the $[111]$ $L$ valley. Labeling of $L$ bands is shown Fig.~\ref{fig:CdTeHgTe}.}
\vspace{0.2cm}
\begin{tabular}{c|| c|c||c | c} \hline\hline
 \multicolumn{1}{c||}{band} &  \multicolumn{2}{c||}{Cd$_{1-x}$Mn$_x$Te}&  \multicolumn{2}{c}{Hg$_{1-x}$Mn$_x$Te} \\
  \hline
      &    TB&       expl&       TB&  expl\\ \hline
$\Gamma_6$&  $\hspace{0.3cm}0.32$ & $\hspace{0.3cm}0.22$ (Ref.\,\onlinecite{Gaj:1979_SSC})  & $\hspace{0.3cm}0.30$& $\hspace{0.3cm}0.40$ (Ref.\,\onlinecite{Furdyna:1988_JAP})\\
$\Gamma_8$&  $-0.78$ &   $-0.88$ (Ref.\,\onlinecite{Gaj:1979_SSC})  &   $-0.74$&   $-0.60$ (Ref.\,\onlinecite{Furdyna:1988_JAP})\\
$\Gamma_7$&  $\hspace{0.3cm}0.24$ & $\hspace{0.3cm}0.29$ (Ref.\,\onlinecite{Twardowski:1980_SSC}) & $\hspace{0.3cm}0.22$ &  \\ \hline
 $L_1$    &  $-0.36$ &            & $-0.22$    & \\
 $L_{2}$  &  $-0.41$ &            & $-0.35$     & \\%
 $L_{3}$  &  $-0.42$ &            & $-0.33$     & \\%
        \hline  
\end{tabular}
 \label{tab:J_cJ_v}
\end{table}

The theory reproduces  the signs and magnitudes of exchange energies at $\Gamma$ properly.
Furthermore, computed data point to a substantial difference in the exchange energy for the conduction band at $L$ compare
to $\Gamma$, disregarded in previous theories of MCD at the $L$ point of the Brillouin zone in Cd$_{1-x}$Mn$_x$Te \cite{Bhattacharjee:1990_PRB}.
A negative sign and a relatively large magnitude of this exchange energy obtained  for Cd$_{1-x}$Mn$_x$Te, $J_c(L_1) = -0.36$, results from a substantial
contribution of the antiferromagnetic kinetic exchange brought about by the proximity of the $L_1$ band to the Mn upper Hubbard band,
the effect described by Eq.~(\ref{eq:sp-d}) and anticipated by {\em ab initio} results shown in Fig.\,\ref{fig:SPINPLITNOSOC_a324}.
The role of kinetic exchange is smaller in Hg$_{1-x}$Mn$_x$Te, where $J_c(L_1) = -0.22$\,eV, because the relativistic mass-velocity term,
large at Hg atoms, shifts downward the conduction band in Hg$_{1-x}$Mn$_x$Te.
This shift makes that the $\Gamma_6$ band is below the $\Gamma_8$ band in Hg$_{1-x}$Mn$_x$Te with low Mn content $x$, so that the material
becomes a topological semimetal.

In the case of the $L_2$ and $L_3$ valence bands, SOC results in spin-momentum locking that diminishes spin-splitting in valleys oblique to the magnetization direction. According to Eqs.\,(\ref{eq: gfact0})-(\ref{eq: gfactpm}), the corresponding geometric factor is given by $\cos[\angle(\mathbf{k}_L, \mathbf{M})]$ for the $L_2$ band; the formulas in the Appendix can be used for deriving
geometry-dependent reduction factors for the two remaining valence bands. Thus, as noted previously \cite{Bhattacharjee:1990_PRB}, a meaningful MCD theory for $E_1$ and $E_1 +\Delta_1$
transitions  requires the evaluation of exchange splittings and transition probabilities for each $L$ valley at a given magnetization direction.

The decomposition of the total Hamiltonian into the exchange-independent and exchange-dependent parts [Eq.\,(\ref{eq:H_TBA})] as well as the use of the Schrieffer-Wolff transformation and atomic values of the potential exchange [Eq.\,(\ref{eq:sp-d})] allows one describing low-energy spin-related effects by a simple Kondo-like Hamiltonian. In order to check a quantitative accuracy of the Schrieffer-Wolff transformation, we have computed the p-d exchange energy N$_0\beta$ by incorporating the Mn d-levels into the TB model employing values of d-level energies and hybridization matrix elements (weighted by x$_{eff}^{1/2}$) collected in Table III.  For x$_{eff}$ = 0.0625, we obtain N$_0\beta$ = -0.76 and -0.72 eV for Cd$_{1-x}$Mn$_x$Te and Hg$_{1-x}$Mn$_x$Te, respectively, the values within 3\% in agreement  with the theoretical data displayed in Table IV, $N_0\beta = -0.78$ and $-0.74$\,eV, respectively.

\subsection{Magnetic circular dichroism for $E_1$ and $E_1 + \Delta_1$ transitions}
\label{sec: MCD}

In order to interpret quantitatively experimental studies of MCD at helium temperatures \cite{Ginter:1983_SSC,Coquillat:1986_SSC,Ando:2011_JAP,Coquillat:1989_PRB}, we have to determine energies and oscillator strengths for optical transitions at the $L$ points of the Brillouin zone. The experimental data were collected  in the Faraday configuration for $\sigma^+$ (electron cyclotron resonance active) and $\sigma^-$ light polarizations, and provided $\Delta E$, that is the difference in spectral positions of edges observed at these two polarizations. This optical exchange splitting $\Delta E$ was found to scale with magnetization, and was independently determined for spectral regions corresponding to $E_1$ and $E_1+\Delta_1$ transitions. In the presence of non-zero magnetization, the four $L$ valleys may not be equivalent and, thus, show different exchange splittings in the presence of SOC. Accordingly, in general, we expect that spectral features may consist of up to 16 excitonic lines in the vicinity of both $E_1$ and $E_1 + \Delta_1$ energy gaps, corresponding to optical transitions $L_2 \rightarrow L_1$ and $L_3 \rightarrow L_1$, respectively. According to the bands' dispersion at the $L$ point shown in Fig.~\ref{fig:CdTeHgTe}, and similarly to the case of other zinc-blende semiconductors \cite{Tanimura:2020_PRB}, $E_1$ and $E_1 + \Delta_1$ optical transitions have a character of saddle-point excitons at low temperatures.

\begin{figure*}[tb]
\includegraphics[width=18cm]{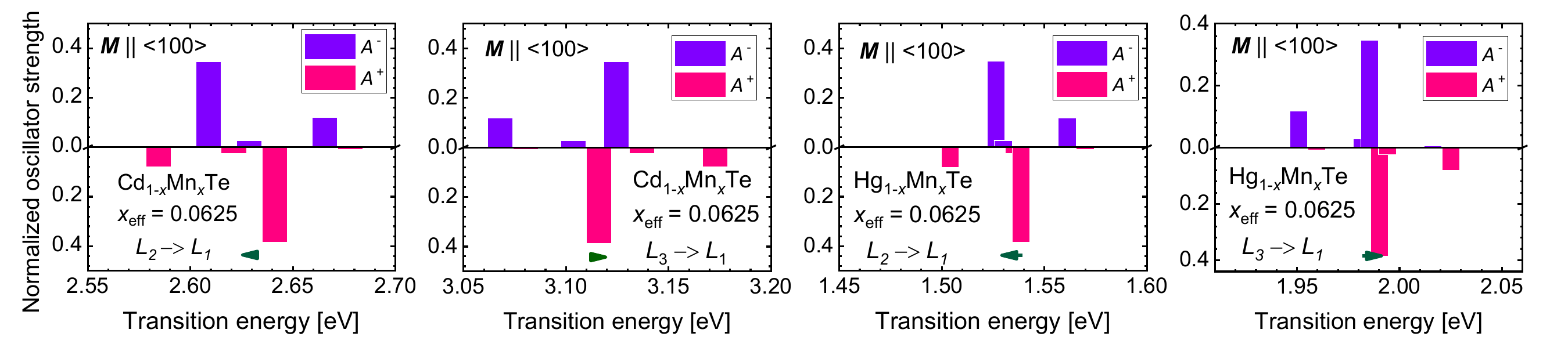}
\includegraphics[width=18cm]{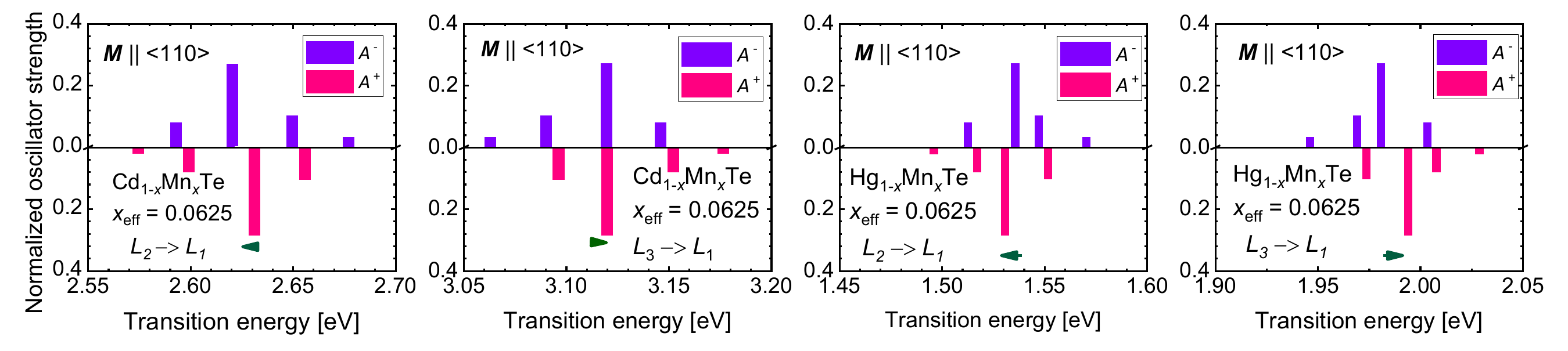}
\includegraphics[width=18cm]{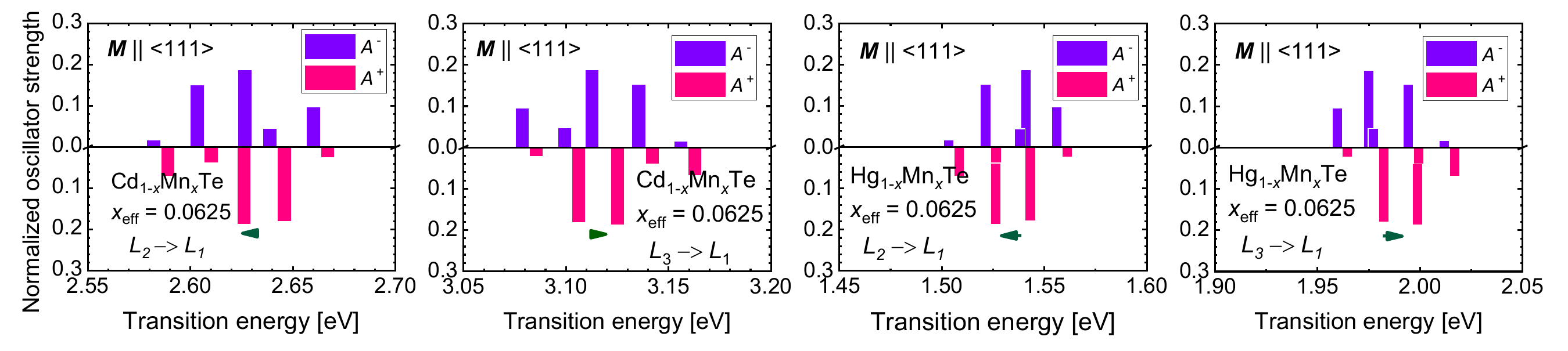}
\caption{Normalized absorption ${\cal A}_{\iota \varphi}^{\pm}$ for $\sigma^+$ and $\sigma^-$ polarizations (positive and negative bars, respectively) for optical transitions in the vicinity of the $E_1$ and $E_1+\Delta_1$ gaps at $L$ points in the Brillouin zone  for Cd$_{1-x}$Mn$_x$Te (left panels) and Hg$_{1-x}$Mn$_x$Te (right panels) with $x_{\text{eff}} =0.0625 $ and fully polarized Mn spins along $\langle100\rangle$,  $\langle110\rangle$, or $\langle111\rangle$ crystal axis that is also the light propagation direction (upper, middle, and bottom panels, respectively). Arrows show the resulting spectral splittings at $L$, $\Delta E$, defined as a distance between the centers of gravity of lines appearing for $\sigma^-$ and $\sigma^+$ polarizations, respectively. In the case of Cd$_{1-x}$Mn$_x$Te, $\Delta E$ at $E_1$ is about 30 times smaller than at $E_0$ for the same Mn magnetization. Note that MCD is usually defined as the difference ${\cal A}^- - {\cal A}^+$.}
\label{fig:MCD}
\end{figure*}

Our tight-binding Hamiltonian (Eq.\,\ref{eq:H_TBA}) allows us to obtain directly expected optical transition energies $E_{\iota\varphi} = E_{\varphi} - E_{\iota}$ between states $\iota$ (initial) and $\varphi$ (final) at critical points of the valence and conduction subbands in a particular $L$-valley at ${\bm k}_L$ and at a given magnetization $\bm{M}$. We assume the transition probabilities at photon energies $h\nu = E_{\iota \varphi}$ are proportional, in the dipole approximation,  to the square
of the interband momentum matrix elements, i.e.,  we have for the oscillator strength $f_{\iota \varphi}^{\pm}$,
\begin{equation}
f_{\iota \varphi}^{\pm} = \frac{2|\langle \iota| p^{\pm}| \varphi\rangle|^2}{m_0 (E_{\varphi} - E_{\iota})},
\end{equation}
where $p^{\pm} = (p_{x} \pm \mbox{i}p_{y})/\sqrt{2}$  for $\sigma^{\pm}$ polarization, respectively, in a right-handed coordinate system $(x, y, z)$ with $z$ being here the optical axis, and $m_0$ is the free electron mass.

The determination of dipole matrix elements is somewhat ambiguous within TBA \cite{Leung:1997_PRB,Lee:2018_PRB}. Similarly, the use of the Peierls-substitution for the determination of the momentum matrix elements in the presence of the vector potential   \cite{Graf:1995_PRB} triggered an extensive debate about the correctness \cite{Pedersen:2001_PRB,Sandu:2005_PRB} and gauge invariance of such a procedure \cite{Boykin:2001_PRB,Foreman:2002_PRB}. Accordingly, optical spectra have been often evaluated within a poorly justified constant momentum matrix element approximation \cite{Czyzyk:1980_pssb} or by treating the momentum matrix elements as a fitting parameter \cite{Leung:1997_PRB}.
As shown in Appendix, we have derived, employing a method of invariants, a $\bm{k} \cdot \bm{p}$ Hamiltonian for the $L$ valleys in zinc-blende semiconductors [equation (\ref{eq: ham})]. A similar formalism was elaborated earlier for the $L$-valleys in Ge and other semiconductors crystallizing in the diamond structure \cite{Li:2012_PRB}. Our general approach, for the $L$ points in zinc-blende semiconductors, takes the inversion asymmetry into account, which results in the presence of terms linear in the ${\bm{k}}$-vector, absent in the diamond structure.
Inherently to this approach, both $f_{\iota \varphi}^{\pm}$ and the effective masses depend on the interband matrix elements $\bm{p}_{vc}$, described by two real material parameters $A$ and $B$. In particular, the inverse effective masses (longitudinal $1/m_l$ and transverse $1/m_t$), characterizing constant energy ellipsoids in the bands denoted as $L_1$ ($L_6$ representation) and $L_2$ ($L_{4,5}$ representation), have contributions proportional to $A^2$ (longitudinal one) and $B^2$ (the transverse one). This is consistent with $|A| \ll |B|$, as $m_l \gg m_t$ according to {\em ab initio} results showed in Fig.~\ref{fig:CdTeHgTe}. Moreover, the magnitudes of $f_{\iota \varphi}$ (for the $E_1$ and $E_1 + \Delta_1$ features) depend mostly on the value of $B$. These imply that the normalized absorbances ${\cal A}_{\iota \varphi}^{\pm}$, i.e., $ f_{\iota \varphi}^{\pm}$  divided by the sum of all $f_{\iota \varphi}^{\pm}$ for either $E_1$ or $E_1 + \Delta_1$ transitions, are immune (within 1\%) to the numerical values of $A$ and $B$ as long as $|A| \lesssim |B|$. Accordingly, theoretical values of MCD can be obtained with no adjustable parameters.

Figure \ref{fig:MCD} shows computed ${\cal A}_{\iota \varphi}^{\pm}$ values for $E_1$ and $E_1+\Delta_1$ transitions involving all four $L$ valleys in Cd$_{1-x}$Mn$_x$Te and Hg$_{1-x}$Mn$_x$Te in the presence of magnetization $\bm{M}$, produced by spin-polarized Mn ions with the concentration $x_{\text{eff}} = 0.0625$ and  $S = 5/2$, parallel to along either $\langle100\rangle$, $\langle110\rangle$, or  $\langle111\rangle$ crystalline axis, which is also the light propagation direction. The TBA has served to determine parameters for the $k \cdot p$ model (including  exchange energies), which in turn has been used to obtain $E_{\iota\varphi}^{\pm}$ and ${\cal A}_{\iota \varphi}^{\pm}$.

In the case of ${\bm M} \| \langle100\rangle$, there are four equivalent valleys tilted by $56.34^{\circ}$ to the light propagation direction, so that we expect that up to four optical transitions in both $E_1$ and $E_1+\Delta_1$ spectral ranges. Since the cleavage plane is $(110)$ for the zinc-blende structure, in the case of bulk single crystal samples \cite{Ginter:1983_SSC,Coquillat:1986_SSC,Coquillat:1989_PRB}, the magnetic field is presumably applied along the $\langle110\rangle$ crystal axis. In this configuration, the ${\bm{k}}_L$ vector of two valleys is perpendicular, and of two others tilted by $35.25^{\circ}$ to the light direction. MCD studies were also carried out on thin films of Cd$_{1-x}$Mn$_x$Te and  Zn$_{1-x}$Mn$_x$Te grown by molecular beam epitaxy on sapphire (0001) substrates, which resulted in (111) oriented polycrystalline samples \cite{Ando:2011_JAP}. In such a case,  light propagates parallel to $L$ valley's longitudinal axis, whereas ${\bm K}_L$ of three other valleys, if effects of strain and piezoelectric fields can be neglected, are tilted by $70.53^{\circ}$ to the light direction.

As seen in Fig.~\ref{fig:MCD}, there are four lines in each polarization for $\bm{M}\| \langle100\rangle$ and $\bm{M}\| \langle110\rangle$. Inspection of the $\bm{M}\| \langle110\rangle$ data shows that the strongest and the weakest feature originate from the two oblique valleys, whereas the two other peaks from the valleys with ${\bm k}_L$ perpendicular to the magnetization and light direction, for which spin-momentum locking by SOC results in vanishing exchange splitting of the valence band states. In the case of $\bm{M}\| \langle111\rangle$, the strongest line comes from the longitudinal valley, ${\bm k}_L \| \langle111\rangle$; the four other lines originate from the three remaining oblique valleys.  For comparison, at $E_0$ in Cd$_{1-x}$Mn$_x$Te, two lines corresponding to the heavy and light excitons are resolved for each polarization. The heavy hole lines are three times stronger, and the energy difference between their positions in the two polarizations is taken as the magnitude  of optical splitting  $\Delta E$ at $\Gamma$, $\Delta E(E_0) = x_{\text{eff}}S(N_0\alpha-N_0\beta)$. This procedure, taking $x_{\text{eff}} = 0.0625$ and the theoretical values of the exchange energies shown in Table~\ref{tab:J_cJ_v}, leads to  $\Delta E(E_0) = 0.17$ and $0.16$\,eV for Cd$_{1-x}$Mn$_x$Te and Hg$_{1-x}$Mn$_x$Te, respectively. According to Fig.~\ref{fig:MCD}, the distances between the strongest lines at $L$ are much smaller, particularly in the case of Cd$_{1-x}$Mn$_x$Te, implying $r = \Delta E(E_0)/\Delta E(E_1) = 38$ and $-17$  for Cd$_{1-x}$Mn$_x$Te and Hg$_{1-x}$Mn$_x$Te, respectively, and ${\bm M}\| [110]$. Actually, taking into account intrinsically large broadening of excitonic transitions at $L$ (Ref.~\cite{Tanimura:2020_PRB}), a better measure of spectral splitting provided by MCD is the distance between centers of gravity of lines appearing in $\sigma^+$ and $\sigma^-$ polarizations, as depicted by arrows in Fig.~\ref{fig:MCD}. In this case,  $r = 33$ and $13$ for the two compounds in question independently of magnetization orientations, the values in a qualitative agreement with the corresponding experimental data $r = 16$ (ref. \cite{Coquillat:1986_SSC,Ando:2011_JAP}) and $5\pm1 $ (ref. \cite{Coquillat:1989_PRB}), respectively. Our theory explains, therefore, a large difference between exchange splittings of optical spectra at $\Gamma$ and $L$ in bulk \cite{Ginter:1983_SSC,Coquillat:1986_SSC,Coquillat:1989_PRB} and epitaxial \cite{Ando:2011_JAP} Cd$_{1-x}$Mn$_x$Te, hitherto  regarded as highly surprising and contradicting theoretical expectations \cite{Ginter:1983_SSC,Coquillat:1986_SSC,Ando:2011_JAP,Bhattacharjee:1990_PRB,Coquillat:1989_PRB}. The developed theory describes also the chemical trend, i.e., a significantly smaller magnitude of $r$ in topological Hg$_{1-x}$Mn$_x$Te \cite{Coquillat:1989_PRB} compared to topologically trivial Cd$_{1-x}$Mn$_x$Te \cite{Ginter:1983_SSC,Coquillat:1986_SSC,Ando:2011_JAP}.  On the other extreme, in the case of light Zn cations,  greater value $r$ was found for bulk Zn$_{1-x}$Mn$_x$Te or even sign reversal of $r$ in the case of epitaxial (111)Zn$_{1-x}$Mn$_x$Te \cite{Ando:2011_JAP}. Furthermore, the computed values of  $\Delta E(E_1+\Delta_1)$ are seen in Fig.~\ref{fig:MCD} to have the same amplitude but the opposite sign compared to $\Delta E(E_1)$, as anticipated earlier \cite{Ginter:1983_SSC,Bhattacharjee:1990_PRB} and observed experimentally for Hg$_{1-x}$Mn$_x$Te \cite{Coquillat:1989_PRB}. Interestingly, no such reversal was fond in the case of epitaxial (111)Cd$_{1-x}$Mn$_x$Te though it appears in epitaxial (111)Zn$_{1-x}$Mn$_x$Te \cite{Ando:2011_JAP}.

\section{Conclusions}

By combining density functional theory with the modified Becke-Johnson exchange-correlation potential, a minimal tight-binding model, and envelope function formalisms at band extrema, we have accurately described experimental energy gaps and exchange splittings of magnetooptical spectra at the $\Gamma$ and $L$ points of the Brillouin zone in Mn-doped topologically trivial  CdTe and topologically non-trivial HgTe with no adjustable or empirical parameters. In particular, according to our insight, a substantial reduction of the exchange-driven splittings at the $L$ points compared to the $\Gamma$ point, since long regarded as challenging \cite{Ginter:1983_SSC,Coquillat:1986_SSC,Ando:2011_JAP,Bhattacharjee:1990_PRB,Coquillat:1989_PRB}, originates from the same sign and similar magnitudes of the exchange energies in the conduction and valence bands at $L$. More specifically, the negative exchange energy of the conduction band at $L$ results from (i) $\bm{k}$-dependent hybridization between band states and Mn open $d$ shells, which leads to the appearance antiferromagnetic kinetic exchange; (ii) $\bm{k}$-dependent changes in the orbital components of the Bloch functions, which affects the relative magnitudes of antiferromagnetic kinetic exchange and ferromagnetic potential exchange, and (iii) the proximity of the conduction band at $L$ to the upper Mn Hubbard band, which enlarges the role of kinetic exchange. This enlargement is more significant in Cd$_{1-x}$Mn$_x$Te compared to Hg$_{1-x}$Mn$_x$Te in which the relativistic mass-velocity term shifts downward $s$-orbitals of Hg contributing to the conduction band energy at $L$. The theory of magnetooptical phenomena also illustrates the interplay of exchange interactions with spin-orbit effects such as spin-momentum locking that diminishes exchange splitting of valence states in $L$-valleys that are oblique to the magnetization direction.

Supplementing previous tight-binding Hamiltonian designed for describing low-energy physics in topological HgTe-CdTe systems \cite{Tarasenko:2015_PRB}, our work provides tight-binding parameters appropriate to investigate phenomena dependent on global properties of the band structure, such as indirect exchange coupling between magnetic ions in magnetic semiconductors. At the same time, the derived $k \cdot p$ Hamiltonian is suitable for modeling phenomena involving $L$ valleys in all compounds with a zinc-blende crystal structure.

\section*{Acknowledgments}

We acknowledge Marcin M. Wysoki{\'n}ski for useful discussions.
The work is supported by the Foundation for Polish Science through the International Research Agendas program co-financed by the
European Union within the Smart Growth Operational Programme.
G. C. acknowledges financial support from the "Fondazione Angelo Della Riccia".
We acknowledge the access to the computing facilities of the Interdisciplinary
Center of Modeling at the University of Warsaw, Grant No.~G73-23 and G75-10.
We acknowledge the CINECA award under the
ISCRA initiatives IsC76 "MEPBI" and IsC81 "DISTANCE"
Grant, for the availability of high-performance computing
resources and support.

\ \\
A.C. and C.\'S. contributed equally to this work.

\appendix*
\section{Determination of the $k \cdot p$ Hamiltonian for the $L$-points in zinc-blende crystals}
\subsection{Construction of the representation for the zinc-blende valence band}

Let $R_0$ be a lattice node (a cation or an anion site) of the zinc-blende lattice and let $G = T_d$ be the corresponding point group, generated by the operations
$\{g_a, g_b\} \subset G$ given in the Cartesian basis as:
\begin{equation}
  g_a = \left( \begin{array}{ccc} 0& 1& 0\\ -1& 0& 0\\ 0& 0& -1 \end{array} \right), \quad
  g_b = \left( \begin{array}{ccc} 0& 0& 1\\ 1& 0& 0\\ 0& 1& 0 \end{array} \right).
  \label{eq: ga, gb}
\end{equation}
A representation $\Delta$ of the space group for the zinc-blende valence band is constructed following Ref.\ \onlinecite{Bradley}:
let $k_L = \frac{\pi}{a_{0}} (1, 1, 1)$ be the $L$-symmetry point of the Brillouin zone, where $a_0$ is the lattice parameter.
The corresponding Hilbert space $H_{k_L}$ of the Bloch states is spanned by the wave-functions $\{ \psi_X, \psi_Y, \psi_Z \} \subset H_{k_L}$, denoted further as kets $X$, $Y$, and $Z$.
The Hilbert space $H_{k_L}$ is the space of the representation $D : G \to B(H_{k_L})$ such that $g_a$ and $g_b$ are represented
by the same matrices (\ref{eq: ga, gb}) in the basis $(X, Y, Z)$.
The primitive translations $t \in \frac{a_0}{2} \{ (0, 1, 1), (1, 0, 1), (1, 1, 0) \}$ act as
\begin{equation}
  \Delta(t) \psi = \exp(-i k_L \cdot t) \psi, \qquad \psi \in H_{k_L}.
\end{equation}
Thus the action of the subgroup
\begin{equation}
  G_{k_L} = \{ g \in G : g k_L \equiv k_L \}
\end{equation}
coincides with the restriction of $D$, $\left. \Delta \right|_{G_{k_L}} = \left. D \right|_{G_{k_L}}$,
while any other operation $g$ moves to another (non-equivalent) $L$-point of the Brillouin zone: $H_{k_L} \ni \psi \mapsto \Delta_g \psi \in H_{g k_L}$.
There are four non-equivalent $L$-points in total,
\begin{eqnarray}
  \lefteqn{G k_L = {}} \nonumber \\
   & & \quad \frac{\pi}{a_{0}} \Bigl\{ (1, 1, 1), (1, -1, -1), (-1, 1, -1), \nonumber \\
   & & \quad\qquad (-1, -1, 1) \Bigr\},
\end{eqnarray}
and the action of $D$ in this set defines the permutation matrices $r_a$ and $r_b$,
\begin{equation}
  r_a = \left( \begin{array}{cccc} 0& 0& 1& 0\\ 1& 0& 0& 0\\ 0& 0& 0& 1\\ 0& 1& 0& 0 \end{array} \right), \qquad
  r_b = \left( \begin{array}{cccc} 1& 0& 0& 0\\ 0& 0& 0& 1\\ 0& 1& 0& 0\\ 0& 0& 1& 0 \end{array} \right).
  \label{eq: ra, rb}
\end{equation}
Therefore,
\begin{equation}
  \Delta(g_a) = r_a \otimes g_a, \qquad
  \Delta(g_b) = r_b \otimes g_b.
\end{equation}
This defines a representation of the space group.

\subsection{Observables}

Accordingly, any observable $\mathcal{O}$ is represented by a $12 \times 12$ matrix, or a $4 \times 4$ block matrix (with $3 \times 3$ blocks).
The action of the primitive translations separates the blocks of $\mathcal{O}$ in the sense that translation-invariant observables are block-diagonal.
Moreover, the requirement of invariance with respect to point operations implies that any such observable is defined by the first block --- the one corresponding to $H_{k_L}$.
For example, scalar observables have the form
\begin{equation}
  \mathcal{O} = \left( \begin{array}{ccc} a_{\mathcal{O}}& b_{\mathcal{O}}& b_{\mathcal{O}}\\
    b_{\mathcal{O}}& a_{\mathcal{O}}& b_{\mathcal{O}}\\
    b_{\mathcal{O}}& b_{\mathcal{O}}& a_{\mathcal{O}}
  \end{array} \right),
\end{equation}
with two real-valued parameters, $(a_{\mathcal{O}}, b_{\mathcal{O}})$.

Vector observables are defined by their Cartesian components, $(\mathcal{V}_x, \mathcal{V}_y, \mathcal{V}_z)$:
\begin{eqnarray}
  \mathcal{V}_x & = & \left( \begin{array}{ccc} a_{\mathcal{V}}& b_{\mathcal{V}}& b_{\mathcal{V}}\\
    c_{\mathcal{V}}& d_{\mathcal{V}}& e_{\mathcal{V}}\\
    c_{\mathcal{V}}& e_{\mathcal{V}}& d_{\mathcal{V}}
  \end{array} \right), \\
  \mathcal{V}_y & = & \left( \begin{array}{ccc} d_{\mathcal{V}}& c_{\mathcal{V}}& e_{\mathcal{V}}\\
    b_{\mathcal{V}}& a_{\mathcal{V}}& b_{\mathcal{V}}\\
    e_{\mathcal{V}}& c_{\mathcal{V}}& d_{\mathcal{V}}
  \end{array} \right), \\
  \mathcal{V}_z & = & \left( \begin{array}{ccc} d_{\mathcal{V}}& e_{\mathcal{V}}& c_{\mathcal{V}}\\
    e_{\mathcal{V}}& d_{\mathcal{V}}& c_{\mathcal{V}}\\
    b_{\mathcal{V}}& b_{\mathcal{V}}& a_{\mathcal{V}}
  \end{array} \right).
  \label{eq: vectobs}
\end{eqnarray}

\subsection{Time-inversion symmetry}

We assume for simplicity that the antiunitary time inversion operator $\hat{\mathcal{T}}$
acts independently on the orbital and spin degrees of freedom,
i.e.\ it is a tensor product
\begin{equation}
 \hat{\mathcal{T}} = \hat{K} \otimes \hat {T}_{1/2}, \qquad
\end{equation}
of some spin-independent part (denoted $\hat K$) and the usual time inversion for spin-$1/2$
particles:
\begin{equation}
 \hat {T}_{1/2} : \psi = \left( \def\arraystretch{1.2} \begin{array}{c} \psi_{\uparrow}\\ \psi_{\downarrow} \end{array} \right)
   \mapsto -i \sigma_y \overline \psi = \left( \def\arraystretch{1.5} \begin{array}{c} -\overline{\psi_{\downarrow}}\\ \overline{\psi_{\uparrow}} \end{array} \right)
\end{equation}
(the arbitrary phase factor being included in $\hat K$).
Therefore, still disregarding the spin for now, we consider an anti-unitary involution $\hat K$,
\begin{equation}
  \hat {K} \left(c \psi\right) = \overline c \hat {K} \psi, \qquad {\hat {K}}^2 \psi = \hat {K} (\hat {K} \psi) = \psi
\end{equation}
(cf.\ the relation $\hat{\mathcal{T}}^2 = -1$ for fermions).
It is defined by the matrix $K_{\alpha\beta} = \left< \psi_\alpha \middle| \hat K \psi_\beta \right>$, ($\alpha, \beta = X, Y, Z$), such that
\begin{equation}
  \hat K \left( \sum_\alpha c_\alpha \psi_\alpha \right) = \sum_{\alpha, \beta} K_{\alpha\beta} \overline{c_\beta} \psi_\alpha.
\end{equation}
Assuming that the representation of the extended symmetry group (space operations and time inversion) is a regular rather than a projective one,
\begin{equation}
  \left( K_{\alpha\beta} \right)_{\alpha, \beta = X, Y, Z} = \left( \begin{array}{ccc} a_T& b_T& b_T\\ b_T& a_T& b_T\\ b_T& b_T& a_T \end{array} \right).
\end{equation}
The pairs $(a_T, b_T)$ such that ${\hat K}^2 = 1$ can be further parameterized by $(u, v)$, $\left| u \right| = \left|v\right| = 1$,
\begin{equation}
  a_T = u \frac{v + 2}{3}, \qquad
  b_T = u \frac{v - 1}{3},
\end{equation}
and the matrix $\left( K_{\alpha\beta} \right)_{\alpha, \beta = X, Y, Z}$ assumes the form
\begin{equation}
  K_{\alpha\beta} = u \left( \delta_{\alpha\beta} + \frac{v - 1}{3} \right).
\end{equation}
Since $u$ can be adjusted by changing the overall phase of the valence-band wave functions,
one lets $u = 1$.

\subsection{Invariant Hamiltonian and the momentum operator}

Accordingly, the invariant Hamiltonian for the valence band at the exact $L$-point reads
\begin{equation}
  H_{inv} = \left( \begin{array}{ccc} a_{H}& b_{H}& b_{H}\\
    b_{H}& a_{H}& b_{H}\\
    b_{H}& b_{H}& a_{H}
  \end{array} \right).
\end{equation}

The construction of a $k \cdot p$ Hamiltonian involves the momentum operator $\hat p$, which we discuss it in detail here.
Momentum is represented as a time-inversion odd, vector operator. The operator equation $\hat K \hat p + \hat p \hat K = 0$
defines a set of homogenous linear equations for the parameters $(a_{p}, b_{p}, c_{p}, d_{p}, e_{p})$ and their complex conjugates,
see Eq.\,(A.8)-(\ref{eq: vectobs}). In order to find the invariants one performs an $L D U$ decomposition
of the coefficients' matrix and investigates the zero entries on the diagonal of $D$. Special combinations
of $(a_T, b_T)$ (or $u$ and $v$) are when a particular entry becomes zero.
Therefore, two cases should be considered:
\begin{enumerate}

  \item If $v = 1$,
\begin{equation}
  \left( K_{\alpha\beta} \right)_{\alpha, \beta = x, y, z} = \left( \begin{array}{ccc} 1& 0& 0\\ 0& 1& 0\\ 0& 0& 1 \end{array} \right).
\end{equation}
and ($P$ real)
\begin{eqnarray}
  \hat{p}_x & = & \frac{m_0 P}{\hbar} \left( \begin{array}{ccc} 0& i& i\\
    -i& 0& 0\\
    -i& 0& 0
  \end{array} \right), \\
  \hat{p}_y & = & \frac{m_0 P}{\hbar} \left( \begin{array}{ccc} 0& -i& 0\\
    i& 0& i\\
    0& -i& 0
  \end{array} \right), \\
  \hat{p}_z & = & \frac{m_0 P}{\hbar} \left( \begin{array}{ccc} 0& 0& -i\\
    0& 0& -i\\
    i& i& 0
  \end{array} \right).
  \label{eq: p1}
\end{eqnarray}

  \item If $v = e^{i \phi} \ne 1$,
\begin{equation}
  \left( K_{\alpha\beta} \right)_{\alpha, \beta = x, y, z} = \frac{1}{3} \left( \begin{array}{ccc} v + 2& v - 1& v - 1\\ v - 1& v + 2& v - 1\\ v - 1& v - 1& v + 2 \end{array} \right),
\end{equation}
and ($P$ real)
\begin{eqnarray}
  \hat{p}_x & = & \frac{2}{3} \sin\left(\frac{\phi}{2}\right) \frac{m_0 P}{\hbar} \left( \begin{array}{ccc} 2& \frac{2 + v}{1 - v}& \frac{2 + v}{1 - v}\\
    -\frac{1 + 2 v}{1 - v}& -1& -1\\
    -\frac{1 + 2 v}{1 - v}& -1& -1
  \end{array} \right), \\
  \hat{p}_y & = & \frac{2}{3} \sin\left(\frac{\phi}{2}\right) \frac{m_0 P}{\hbar} \left( \begin{array}{ccc} -1& -\frac{1 + 2 v}{1 - v}& -1\\
    \frac{2 + v}{1 - v}& 2& \frac{2 + v}{1 - v}\\
    -1& -\frac{1 + 2 v}{1 - v}& -1
  \end{array} \right), \\
  \hat{p}_z & = & \frac{2}{3} \sin\left(\frac{\phi}{2}\right) \frac{m_0 P}{\hbar} \left( \begin{array}{ccc} -1& -1& -\frac{1 + 2 v}{1 - v}\\
    -1& -1& -\frac{1 + 2 v}{1 - v}\\
    \frac{2 + v}{1 - v}& \frac{2 + v}{1 - v}& 2
  \end{array} \right).
  \label{eq: p2}
\end{eqnarray}

\end{enumerate}

\subsection{The $k \cdot p$ Hamiltonian}

The standard derivation of $k \cdot p$ Hamiltonians involves perturbation theory and an expansion in the basis of the high-symmetry-point Bloch functions.
Since in a typical semiconductor the result is dominated by terms due to virtual transitions to the conduction band, rather then giving the general form of the $k \cdot p$ Hamiltonian,
here we write out an $8 \times 8$ matrix (including the conduction band explicitly rather than via perturbation theory, and a spin-orbit term) which preserves
the most important features of such systems. In the basis of $(S, X, Y, Z)$ states our Hamiltonian ${\cal H}$ reads:
\begin{eqnarray}
  {\cal H} & = & \left( \begin{array}{cc} E_{c} + \frac{\hbar}{m_0} \mathbf{k} \cdot \hat{\mathbf{p}}_c& \frac{\hbar}{m_0} \mathbf{k} \cdot \hat{\mathbf{p}}_{vc} \\ \frac{\hbar}{m_0} \mathbf{k} \cdot \hat {\mathbf{p}}_{vc}^{\dagger}& H_{inv} + \frac{\hbar}{m_0} \mathbf{k} \cdot \hat{\mathbf{p}} \end{array} \right) \nonumber \\
  & & \qquad {}  + \frac{1}{3} \lambda_{SO} (I'_x \sigma_x + I'_y \sigma_y  + I'_z \sigma_z),
\end{eqnarray}
where $\hat{\mathbf{p}}_c$ is the momentum operator in the conduction band (it is given
in the next subsection in coordinates appropriate for the $L$-point).
The valence band orbital momentum matrices $I'_\alpha$ are defined as usual by the Levi-Civita symbol $(I_\alpha)_{\beta\gamma} = -i \varepsilon_{\alpha\beta\gamma}$, $\alpha, \beta, \gamma = x, y, z$
by padding with zeros to a $4 \times 4$ matrix, whereas $\sigma_\alpha$ are the Pauli matrices corresponding to the spin degree of freedom.

The interband matrices of momentum have the general form ($a_{pvc}, b_{pvc}$ complex):
\begin{eqnarray}
  p_{vc,x} & = & \left( \begin{array}{ccc} a_{pvc}& b_{pvc}& b_{pvc} \end{array} \right), \\
  p_{vc,y} & = & \left( \begin{array}{ccc} b_{pvc}& a_{pvc}& b_{pvc} \end{array} \right), \\
  p_{vc,z} & = & \left( \begin{array}{ccc} b_{pvc}& b_{pvc}& a_{pvc} \end{array} \right).
\end{eqnarray}
We require again that $\hat{\mathbf{p}}_{vc}$ is odd w.r.t.\ the time inversion, where the time inversion in the conduction band
is defined as $\hat{\mathcal{T}} \left| S \uparrow \right> = \left| S \downarrow \right>$, $\hat{\mathcal{T}} \left| S \downarrow \right> = -\left| S \uparrow \right>$.
%
%
%
%
There exists two such invariants, therefore we introduce a complex parameter $z$, and
\begin{equation}
  a_{pvc} = \frac{m_0}{\hbar} \frac{(1 + 2 v) z + 3 \overline{z}}{1 - v}, \qquad b_{pvc} = \frac{m_0}{\hbar} z,
\end{equation}
unless $v = 1$, when
\begin{equation}
  a_{pvc} = \frac{i m_0}{\hbar} a, \qquad b_{pvc} = \frac{i m_0}{\hbar} b,
\end{equation}
$a$ and $b$ being two real parameters.

\subsection{Change of coordinates}

In order to diagonalize $H_{inv}$, one transforms the basis of the Hilbert space $H_{k_L}$ according to ($\omega = e^{2 \pi i/3}$)
\begin{equation}
  U_\omega = \frac{1}{\sqrt{3}} \left( \begin{array}{ccc} 1& \omega& \omega^2\\ 1& \omega^2& \omega\\ 1& 1& 1 \end{array} \right).
\end{equation}
In the new basis $H_{inv}$ is diagonal, with eigenvalues $E_{v1} = a_H + 2 b_H$ (non-degenerate) and $E_{v2} = a_H - b_H$ (two-fold degenerate).
$\hat K$ swaps the two energy-degenerate eigenstates, and is $v$ on the non-degenerate eigenstate. By an appropriate
adjustment of the phase of the corresponding basis vector $v$ can be set to $v = 1$. Let us assume this case.

Furthermore, an appropriate coordinate system with $z'$ along $[111]$ is introduced by the rotation $R_{[111]}$,
\begin{equation}
  R_{[111]} = \left( \begin{array}{ccc} 1/\sqrt{2}& 1/\sqrt{6}& 1/\sqrt{3}\\ -1/\sqrt{2}& 1/\sqrt{6}& 1/\sqrt{3}\\ 0& -2/\sqrt{6}& 1/\sqrt{3} \end{array} \right),
\end{equation}
which allows to define two parameters $A = -\frac{i \hbar}{m_0} (a_{pvc} + 2 b_{pvc})$ and $B = -\frac{i \hbar}{m_0} (a_{pvc} - b_{pvc})$. 
As discussed above, we assume that the time inversion parameters $u = v = 1$, which implies that $A$ and $B$ are real.
These interband momentum matrix elements determine the band dispersion in the vicinity of the $L$-point,
in the longitudinal and transverse directions respectively. Position operator $\hat{\mathbf{r}}$
is given by matrices of the same form, but then the matrix elements $(a_{rvc}, b_{rvc})$ are real rather than purely imaginary.

The rotation of the real-space coordinates implies a rotation in the space of the spin degrees of freedom. We implement it with a unitary transformation $U_c$,
\begin{equation}
  U_c = \frac{1}{\sqrt{6 - 2 \sqrt{3}}} \left( \begin{array}{cc} 1 - i& 1 - \sqrt{3}\\ -1 + \sqrt{3}& 1 + i \end{array} \right),
\end{equation}
where the group generators $g_a$ and $g_b$ are represented as
\begin{eqnarray}
  g_a & \mapsto & \frac{1}{\sqrt{2}} \left( \begin{array}{cc} 1 - i& 0\\ 0& 1 + i \end{array} \right), \\
  g_b & \mapsto & \frac{1}{2} \left( \begin{array}{cc} 1 - i& -1 - i\\ 1 - i& 1 + i \end{array} \right).
\end{eqnarray}
The transformed Pauli matrices read:
\begin{eqnarray}
  \sigma_{x'}' & = & -\frac{1}{\sqrt{2}} \left( \begin{array}{cc} 0& 1 - i\\ 1 + i& 0\end{array} \right), \\
  \sigma_{y'}' & = & \frac{1}{\sqrt{2}} \left( \begin{array}{cc} 0& 1 + i\\ 1 - i& 0\end{array} \right), \\
  \sigma_{z'}' & = & \left( \begin{array}{cc} 1& 0\\ 0& -1\end{array} \right).
\end{eqnarray}
With this notation, the conduction-band momentum operator takes the form
($Q / \hbar$ is the transverse velocity in the conduction band):
\begin{equation}
  p_{c,x'}' = \frac{m_0}{\hbar} Q \sigma_{y'}', \qquad
  p_{c,y'}' = -\frac{m_0}{\hbar} Q \sigma_{x'}', \qquad
  p_{c,z'}' = 0.
\end{equation}
In the new coordinates, the valence-band spin-orbit interaction term (in a basis in which the spin-up states
precede those with spin-down) takes the form
\begin{equation}
  {\cal H}_{SO} = \frac{1}{3} \lambda_{SO} \left( \begin{array}{cccccc}
    0& 0& 0& 0& 1 + i& 0\\
    0& 1& 0& 0& 0& 0\\
    0& 0& -1& -1 - i& 0& 0\\
    0& 0& -1 + i& 0& 0& 0\\
    1 - i& 0& 0& 0& -1& 0\\
    0& 0& 0& 0& 0& 1
  \end{array} \right).
\end{equation}
More generally, the diagonal entries in this matrix can have a (longitudinal) coefficient independent
from the coefficient of the off-diagonal entries (a transverse one).

The Hamiltonian 
is displayed in Eq.\ (\ref{eq: ham}). In addition to the energies $E_c$, $E_{v1}$, $E_{v2}$, and $\lambda_{SO}$
[all determined by the empirical band energies at $L$ according to (\ref{eq:eigenvalues})],
it is characterized by the four momentum matrix elements $Q$, $P$, $A$, and $B$.
\begin{widetext}
\begin{sidewaysfigure}
\begin{equation}
  {\cal H} = \left( \begin{array}{cccccccc}
    E_c& Q \left( \frac{1 + i}{\sqrt{2}} k_{x'} + \frac{1 - i}{\sqrt{2}} k_{y'} \right)& i A k_{z'}& -\frac{B (k_{x'} + i k_{y'})}{\sqrt{2}}& \frac{B (k_{x'} - i k_{y'})}{\sqrt{2}}& 0& 0& 0\\
    Q \left( \frac{1 - i}{\sqrt{2}} k_{x'} + \frac{1 + i}{\sqrt{2}} k_{y'} \right)& E_c& 0& 0& 0& i A k_{z'}& -\frac{B (k_{x'} + i k_{y'})}{\sqrt{2}}& \frac{B (k_{x'} - i k_{y'})}{\sqrt{2}}\\
    -i A k_{z'}& 0& E_{v1}& \sqrt{\frac{3}{2}} P (k_{x'} + i k_{y'})& -\sqrt{\frac{3}{2}} P (k_{x'} - i k_{y'})& 0& \frac{1 + i}{\sqrt{3}} \lambda_{SO}& 0\\
    -\frac{B (k_{x'} - i k_{y'})}{\sqrt{2}}& 0& \sqrt{\frac{3}{2}} P (k_{x'} - i k_{y'})& E_{v2} + \frac{\lambda_{SO}}{3}& 0& 0& 0& 0\\
    \frac{B (k_{x'} + i k_{y'})}{\sqrt{2}}& 0& -\sqrt{\frac{3}{2}} P (k_{x'} + i k_{y'})& 0& E_{v2} - \frac{\lambda_{SO}}{3}& -\frac{1 + i}{\sqrt{3}} \lambda_{SO}& 0& 0\\
    0& -i A k_{z'}& 0& 0& -\frac{1 - i}{3} \lambda_{SO}& E_{v1}& \sqrt{\frac{3}{2}} P (k_{x'} + i k_{y'})& -\sqrt{\frac{3}{2}} P (k_{x'} - i k_{y'})\\
    0& -\frac{B (k_{x'} - i k_{y'})}{\sqrt{2}}& \frac{1 - i}{3} \lambda_{SO}& 0& 0& \sqrt{\frac{3}{2}} P (k_{x'} - i k_{y'})& E_{v2} - \frac{\lambda_{SO}}{3}& 0\\
    0& \frac{B (k_{x'} + i k_{y'})}{\sqrt{2}}& 0& 0& 0& -\sqrt{\frac{3}{2}} P (k_{x'} + i k_{y'})& 0& E_{v2} + \frac{\lambda_{SO}}{3}
  \end{array} \right).
  \label{eq: ham}
\end{equation}
\end{sidewaysfigure}
\end{widetext}
In the same basis, the generators of the $C_{3v}$ group are
\begin{eqnarray}
  c_{[111]} & = & \mathop{\mathrm{diag}}\Bigl(\frac{1 - i \sqrt{3}}{2}, \frac{1 + i \sqrt{3}}{2}, \frac{1 - i \sqrt{3}}{2}, -1, \Bigr. \nonumber \\
   & & \qquad \Bigl. \frac{1 + i \sqrt{3}}{2}, \frac{1 + i \sqrt{3}}{2}, \frac{1 - i \sqrt{3}}{2}, -1 \Bigr)
\end{eqnarray}
(the threefold rotation), and 
\begin{equation}
  s_{(1\overline{1}0)} = \left( \begin{array}{cccccccc}
    0& \frac{1 + i}{\sqrt{2}}& 0& 0& 0& 0& 0& 0\\
    -\frac{1 - i}{\sqrt{2}}& 0& 0& 0& 0& 0& 0& 0\\
    0& 0& 0& 0& 0& \frac{1 + i}{\sqrt{2}}& 0& 0\\
    0& 0& 0& 0& 0& 0& 0& \frac{1 + i}{\sqrt{2}}\\
    0& 0& 0& 0& 0& 0& \frac{1 + i}{\sqrt{2}}& 0\\
    0& 0& -\frac{1 - i}{\sqrt{2}}& 0& 0& 0& 0& 0\\
    0& 0& 0& 0& -\frac{1 - i}{\sqrt{2}}& 0& 0& 0\\
    0& 0& 0& -\frac{1 - i}{\sqrt{2}}& 0& 0& 0& 0
  \end{array} \right).
  \label{eq: mirror}
\end{equation}
(the reflection). The eigenvalues of $c_{[111]}$ equal to $-1$ correspond to the $L_{4,5}$ representation, the remaining ones to $L_6$.

The eigenvalues of $H$ at $k = 0$ (the exact $L$-point) are $E_c$ and:
\begin{eqnarray} \label{eq:eigenvalues}
E_0 & = & E_{v2} + \frac{1}{3} \lambda_{SO} \nonumber \\
E_{\pm} & = & \frac{1}{6} \biggl[ 3 E_{v1} + 3 E_{v2} - \lambda_{SO} \nonumber \\
  &  \bigl. \pm & \sqrt{9 (E_{v1} - E_{v2})^2 + 6 (E_{v1} - E_{v2}) \lambda_{SO} + 9 \lambda_{SO}^2} \biggr]
\end{eqnarray}
and are (Kramers) twofold-degenerate. The corresponding eigenvectors are
\begin{eqnarray}
  \psi_{c\uparrow} & = & \left| 1 \right>, \\
  \psi_{c\downarrow} & = & \left| 2 \right>, \\
  \psi_{0\uparrow} & = & \left| 4 \right>, \\
  \psi_{0\downarrow} & = & \left| 8 \right>, \\
  \psi_{{\pm}\uparrow} & = & \frac{1}{n_{\pm}} \left[ c_{\pm} \left| 3 \right> + (1 - i) d_{\pm} \left| 7 \right> \right], \\
  \psi_{{\pm}\downarrow} & = & \frac{1}{n_{\pm}} \left[ -(1 + i)  d_{\pm} \left| 5 \right> + c_{\pm} \left| 6 \right> \right]
\end{eqnarray}
($n_{\pm} > 0$ is the normalization) with
\begin{eqnarray}
  \lefteqn{c_{\pm} = 3 E_{v1} - 3 E_{v2} + \lambda_{SO}} \nonumber \\
  & & \pm \sqrt{9 (E_{v1} - E_{v2})^2 + 6 (E_{v1} - E_{v2}) \lambda_{SO} + 9 \lambda_{SO}^2}, \\
  \lefteqn{d_{\pm} = 2 \lambda_{SO}.}
\end{eqnarray}
In the basis of eigenstates, the threefold rotation $c_{[111]}$ is diagonal, with eigenvalues
\begin{eqnarray}
  \lefteqn{\mathop{\mathrm{diag}}\Bigl( \frac{1 - i \sqrt{3}}{2}, \frac{1 + i \sqrt{3}}{2}, -1, -1, \Bigr.} \nonumber \\
  & & \qquad \Bigl. \frac{1 - i \sqrt{3}}{2},
    \frac{1 + i \sqrt{3}}{2}, \frac{1 - i \sqrt{3}}{2}, \frac{1 + i \sqrt{3}}{2} \Bigr)
\end{eqnarray}
and the reflection $s_{(1\overline{1}0)}$ is block-diagonal, with each block of the same form:
\begin{equation}
  \frac{1}{\sqrt{2}} \left( \begin{array}{cc} 0& 1 + i\\ -1 + i& 0\end{array} \right).
\end{equation}
The time inversions acts as (now the spin directions are given w.r.t.\ the $[111]$ axis)
\begin{eqnarray}
  \hat{\mathcal{T}} \left( \psi_{c\uparrow} \right) & = & \overline{\psi_{c\downarrow}}, \\
  \hat{\mathcal{T}} \left( \psi_{c\downarrow} \right) & = & \overline{-\psi_{c\uparrow}}, \\
  \hat{\mathcal{T}} \left( \psi_{0\uparrow} \right) & = & \overline{\psi_{0\downarrow}}, \\
  \hat{\mathcal{T}} \left( \psi_{0\downarrow} \right) & = & \overline{-\psi_{0\uparrow}}, \\
  \hat{\mathcal{T}} \left( \psi_{\pm\uparrow} \right) & = & \overline{\psi_{\pm\downarrow}}, \\
  \hat{\mathcal{T}} \left( \psi_{\pm\downarrow} \right) & = & \overline{-\psi_{\pm\uparrow}}.
\end{eqnarray}

Furthermore, the restrictions of the spin operator $(s_{x'}', s_{y'}', s_{z'}')$ to the eigenspaces take the form:
\begin{equation}
  \left. s_{x',y',z'}' \right|_{E = E_{c}} = \frac{\sigma_{x',y',z'}'}{2},
\end{equation}
\begin{equation}
  \left. s_{x',y'}' \right|_{E = E_{0}} = 0, \qquad
  \left. s_{z'}' \right|_{E = E_{0}} = \frac{\sigma_{z'}'}{2},
  \label{eq: gfact0}
\end{equation}
\begin{eqnarray}
  \lefteqn{\left. s_{x',y'}' \right|_{E = E_{\pm}} = {}} \nonumber \\
   & & {} \pm \frac{3 E_{\pm} - 3 E_{v2} + \lambda_{SO}}{\sqrt{9 (E_{v1} - E_{v2})^2 + 6 (E_{v1} - E_{v2}) \lambda_{SO} + 9 \lambda_{SO}^2}} \nonumber \\ & & \qquad \qquad \qquad \qquad \qquad \qquad \qquad \qquad {} \times  \frac{\sigma_{x',y'}'}{2}, \\
  \lefteqn{\left. s_{z'}' \right|_{E = E_{\pm}} = {}} \nonumber \\
   & & {} \pm \frac{3 E_{v1} - 3 E_{v2} + \lambda_{SO}}{\sqrt{9 (E_{v1} - E_{v2})^2 + 6 (E_{v1} - E_{v2}) \lambda_{SO} + 9 \lambda_{SO}^2}} \nonumber \\ & & \qquad \qquad \qquad \qquad \qquad \qquad \qquad \qquad {} \times \frac{\sigma_{z'}'}{2}.
  \label{eq: gfactpm}
\end{eqnarray}


\subsection{Velocities and effective masses}

The bands can be classified as those with linear ($\left. \partial E / \partial k \right|_{k = 0} \ne 0$) and parabolic ($\left. \partial E / \partial k \right|_{k = 0} = 0$) transverse (in the $k_{x'}, k_{y'}$ plane) dispersion.
Generally, for a twofold degenerate band, the velocities are given in terms of the characteristic polynomial $\chi = \chi(k, E)$ as
\begin{equation}
  \hbar v = -\frac{\left( \frac{\partial^2 \chi(k, E)}{\partial k \, \partial E} \right)}{\left( \frac{\partial^2 \chi(k, E)}{\partial E^2} \right)} \pm \sqrt{\left[ \frac{\left( \frac{\partial^2 \chi(k, E)}{\partial k \, \partial E} \right)}{\left( \frac{\partial^2 \chi(k, E)}{\partial E^2} \right)} \right]^2 - \frac{\left( \frac{\partial^2 \chi(k, E)}{\partial k^2} \right)}{\left( \frac{\partial^2 \chi(k, E)}{\partial E^2} \right)}}.
\end{equation}
In the present case, $\chi(k, E) = \chi(-k, E)$ due to Kramers degeneracy and one writes $\chi = \chi(k^2, E)$. For the same reason, $\partial \chi(0, E)/\partial E = 0$. Then,
in the former class (the conduction band and the $E_{\pm}$ eigenspaces), the velocity can be calculated by evaluating the following expression at $k = 0$:
\begin{equation}
  \frac{1}{\hbar} \left| \frac{\partial E}{\partial k} \right| = \frac{1}{\hbar} \sqrt{\frac{-2 \left( \frac{\partial \chi(k^2, E)}{\partial (k^2)} \right)}{\left(\frac{\partial^2 \chi(k^2, E)}{\partial E^2} \right)}},
\end{equation}
or by diagonalization of the velocity operator, $\frac{1}{\hbar} \left. \partial H(\mathbf{k})/\partial \mathbf{k} \right|_{E = E_{\pm}}$ in each energy eigenspace.
The result for the transverse velocity in the ${\pm}$ band (disregarding sign) is
\begin{equation}
  v_{\pm} = \frac{1}{\hbar} \frac{2 P \lambda_{SO}}{\sqrt{3 (E_{v1} - E_{v2})^2 + 2 (E_{v1} - E_{v2}) \lambda_{SO} + 3 \lambda_{SO}^2}}.
\end{equation}
Therefore, the dispersion $E(k)$ can be approximated as
\begin{equation}
  E(k) = E(0) \pm \hbar v_t \sqrt{k_{x'}^2 + k_{y'}^2} + \frac{\hbar^2}{2 m_t}(k_{x'}^2 + k_{y'}^2) + \frac{\hbar^2}{2 m_l} k_{z'}^2,
\end{equation}
and the masses $(m_i)_{i = t, l}$ are given by
\begin{equation}
  \frac{\hbar^2}{2m} = \frac{\frac{1}{6} \frac{\partial^2 \chi(k, E)}{\partial k^2} \frac{\partial^3 \chi(k, E)}{\partial E^3}
    - \frac{1}{2} \frac{\partial^3 \chi(k, E)}{\partial k^2 \, \partial E} \frac{\partial^2 \chi(k, E)}{\partial E^2}}{\left[ \frac{\partial^2 \chi(k, E)}{\partial E^2} \right]^2}
\end{equation}
\begin{equation}
  \frac{\hbar^2}{2m} = \frac{\frac{1}{3} \frac{\partial \chi(k^2, E)}{\partial (k^2)} \frac{\partial^3 \chi(k^2, E)}{\partial E^3}
    - \frac{\partial^2 \chi(k^2, E)}{\partial (k^2) \, \partial E} \frac{\partial^2 \chi(k^2, E)}{\partial E^2}}{\left[ \frac{\partial^2 \chi(k^2, E)}{\partial E^2} \right]^2}
\end{equation}
Explicitly, each $\hbar^2/2 m_i$ has contributions from
\begin{enumerate}
  \item the value of the second derivative of the Hamiltonian, $\frac{1}{2} \partial^2 {\cal H}(k)/\partial k^2$, in the energy eigenspace;
  \item the second order off-diagonal perturbations, $\frac{1}{2} \sum_{j \ne i} \mathop{\mathrm{Tr}} \left[ \left( {\cal H}'_{ij} {\cal H}'_{ij}\right)^{\dagger} \right]/(E_i - E_j)$,
    where ${\cal H}'_{ij}$ denotes the $2 \times 2$ off-diagonal block of $\partial \mathcal{H}(k)/\partial k$ corresponding to bands $(i, j)$.
\end{enumerate}

\bibliography{Lsplitting9Feb2021}

\end{document}